\documentclass[preprint2]{aastex}

\usepackage{rotating,lscape,natbib}

\shorttitle{Rotation velocities for M-dwarfs}
\shortauthors{Jenkins et al.}

\begin{document}


\title{Rotation Velocities for M-dwarfs\footnote{Based on observations obtained with the Hobby-Eberly Telescope, which is a joint project of the University of Texas at Austin, the Pennsylvania State University, Stanford University, Ludwig-Maximilians-Universit\"{a}t M\"{u}nchen, and Georg-August-Universit\"{a}t G\"{o}ttingen.}}


\author{J S Jenkins$^{1,2}$, L W Ramsey$^{1}$, H R A Jones$^{3}$, Y Pavlenko$^{3}$, J Gallardo$^{2}$, J R Barnes$^{3}$ and D J Pinfield$^{3}$}
\affil{$^1$Department of Astronomy and Astrophysics, Pennsylvania State University, University Park, PA16802\\
       $^2$Department of Astronomy, Universidad de Chile, Casilla Postal 36D, Santiago, Chile\\
       $^3$Center for Astrophysics, University of Hertfordshire, College Lane Campus, Hatfield, Hertfordshire, UK, AL10 9AB}
\email{jjenkins@astro.psu.edu}


\begin{abstract}

We present spectroscopic rotation velocities ($v$~sin~$i$) for 56 M dwarf stars using high resolution HET HRS 
red spectroscopy.  In addition we have also determined photometric 
effective temperatures, masses and metallicities ([Fe/H]) for some stars observed here and in the literature where we could acquire accurate parallax measurements and relevant photometry.  
We have increased the number of known $v$~sin~$i$s for mid M stars by
around 80\% and can confirm a weakly increasing rotation velocity with decreasing effective 
temperature.  Our sample of $v$~sin~$i$s peak at low velocities ($\sim$3~km~s$^{-1}$).  We find a change in the rotational velocity distribution between early M and 
late M stars, which is likely due to the changing field topology between partially and fully convective stars.  There is also a possible further change in the rotational distribution towards 
the late M dwarfs where dust begins to play a role in the stellar atmospheres.  We also link $v$~sin~$i$ to age and show how it can be used to provide mid-M star age limits.

When all literature velocities for M dwarfs are added to our sample there are 198 with $v$~sin~$i$~$\le$~10~km~s$^{-1}$ and 124 in the mid-to-late M star regime (M3.0-M9.5) where 
measuring precision optical radial-velocities is difficult.  In addition we also search the spectra for any 
significant H$\alpha$ emission or absorption.  43\% were found to exhibit such emission and could represent young, active objects with high levels of radial-velocity 
noise.  We acquired two epochs of spectra for the star GJ1253 spread by almost one month and the H$\alpha$ profile changed from showing no clear signs of emission, to exhibiting a clear 
emission peak.  Four stars in our sample appear to be low-mass binaries (GJ1080, GJ3129, Gl802 and LHS3080), with both GJ3129 and Gl802 exhibiting double H$\alpha$ emission 
features.  The tables presented here will aid any future M star planet search target selection to extract stars that will exhibit low radial-velocity jitter. 

\end{abstract}





\section{Introduction}

In the past 10 years radial-velocity measurements in the optical have made great strides by utilising a number of techniques and methodologies to generate 
precisions of 3~m~s$^{-1}$ in the long term (e.g. \citealp{butler06}) and sub-m~s$^{-1}$ in the short term (e.g. \citealp{bouchy05}).  Use of an iodine cell (e.g. 
\citealp{marcy92}) or ThAr gas lamp (e.g. \citealp{pepe00}) have 
allowed detection of around 300 extrasolar planets (exoplanets), with that number increasing each month (see http://exoplanet.eu/).  A large parameter space has been studied 
but since these observations are limited to the optical regime, where M dwarfs are intrinsically faint, a vast amount of stars are left unobserved, particularly the M star population 
that constitutes the bulk of 
stars in the local galactic neighbourhood. Radial-velocity studies of early M dwarfs have detected planets well into the terrestrial-mass 
regime, down below 10M$_{\oplus}$ (e.g. \citealp{rivera05}; \citealp{udry07}; \citealp{mayor09}).  There is considerably improved mass contrast with such stars (amplitude of a given mass planet 
$\propto$ M$_{\rm{star}}$$^{0.5}$) and obtaining precision radial-velocities of these objects may enable Earth-mass planets to be detected in their habitable zones.

The largest uncertainty associated with precision radial-velocity measurements is that of stellar activity.  The association between stellar activity and rotation 
velocity ($v$~sin~$i$) is well established (\citealp{noyes}) and measuring this parameter is an excellent proxy of the level of activity.  
For a fixed resolution, S/N and calibration method, stars with higher rotational velocities have radial-velocity measurements with lower precision 
since the larger rotation serves to wash out the spectral features (\citealp{bouchy01}).  Coupled to 
this, the combination of activity and rotation can cause false positives to appear in the data (e.g. \citealp{henry02}) but in general it serves to 
increase the level of jitter (\citealp{wright05}).  However, the rotation-activity connection has been shown to saturate quickly around M stars, 
occuring at $\sim$5~km~s$^{-1}$ for M2 objects (\citealp{patten96}).  Since we are probing later M stars ($\ge$M3), where observations of various activity 
indicators have shown that the percentage of active stars will increase dramatically, up to $\sim$100\% at M7 (\citealp{fleming00}; \citealp{mohanty02}),
we are focused on selecting the slowest rotators within each spectral bin, in order to select the narrowest line profiles.  An increase in radial-velocity 
uncertainty of around 3-4 times was found by Bouchy et al. when increasing the $v$~sin~$i$ of stellar models from 1-10~km~s$^{-1}$.  The larger 
uncertainty arises due to the loss of spectral information in the stars with higher 
$v$~sin~$i$  since the blending factor is increased within the stellar forrest, to the detrement of the radial-velocity information.  

Based on the above description any future precision radial-velocity planet search survey targeting cool M stars should have a fairly strict selection based on the rotation velocity 
of their sample stars, particularly when targeting Earth-mass planets in the habitable zones.  For instance, a search employed using a PRVS-like instrument (\citealp{jones08}; 
\citealp{ramsey08}) would focus on selecting the brightest M stars 
with rotational velocities in the sub-10~km~s$^{-1}$ regime, which would allow high S/N spectra to be acquired in the shortest possible observational times, gaining in the number of M stars 
observed in a single observing run, and also allowing the highest precisions to be reached to detect the lowest mass rocky planets, in particular Earth-like planets in their habitable 
zones.  We do note that such a 10~km~s$^{-1}$ upper limit should probably be raised when moving into the ultracool/brown dwarf regime to increase the number of such objects on any 
planet search sample since previous studies have shown that rotation velocities appear to systematically increase (e.g. \citealp{mohanty03}; \citealp{reiners08}).

\section{Observations \& Reduction}

All observations in this study were made over the period 2006~December~25$-$2007~November~04 utilising the queue scheduling mode at the 9.2m Hobby-Eberly Telescope (HET; 
\citealp{ramsey98}) at McDonald Observatory in Texas.  Using the High Resolution Spectrograph (HRS; \citealp{tull98}) red chip ($\sim$8100-9900\AA) operating at a resolution of 
\emph{R}$\sim$37,000 and spatial binning of 2x2, 53 objects were observed, ranging in spectral type from M3V to M6.5V.  The HRS observing mode 
employed the use of two 3$''$ fibers since the objects were so faint, with one on object and one on sky.  Bright telluric standards taken from the HET list of 
rapidly rotating B stars were also observed on most of the nights, which allowed us to characterize the level of telluric contamination in the data.  
Since even close by mid-to-late M stars are relatively faint, our 
HRS integration times ranged in duration from 5-25~minutes and for some of the fainter objects multiple exposures were taken and combined to generate a high S/N spectrum.  
However, most of the $v$~sin~$i$ measurements, when comparing the individual frames to the combined frame, were within $\pm$0.5~km~s$^{-1}$ since the deconvolution method we employ boosts the S/N in 
the final line profile used in the measurements.

\begin{figure}
\vspace{5.0cm}
\hspace{-4.0cm}
\includegraphics{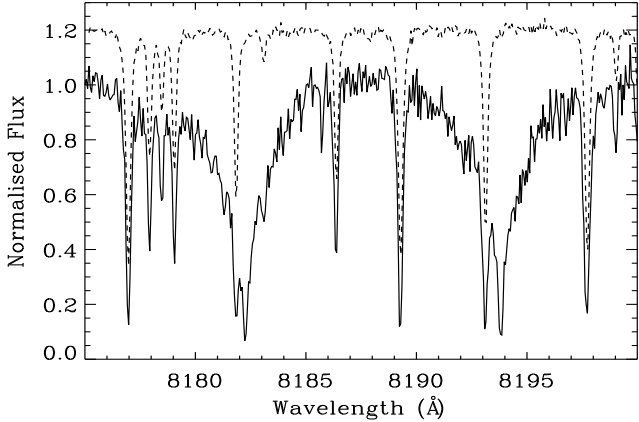}
\vspace{0.5cm}
\caption{The normalized HRS spectra for the M4.5 dwarf G~121-028 around the sodium doublet (8182-8194\AA) is shown by the solid line.  The dashed line represents the telluric standard 
HD1839 and has been offset for clarity.  Note the blending between the strong sodium lines and the tellurics in the M dwarf spectra.}
\label{spectra}
\end{figure}

The reduction of all data was performed using \sc ccdpack \rm and \sc starlink \rm techniques.  First the bias and overscan signals were removed and the overscan region 
was clipped off each image.  A master flatfield was created by median filtering all flats taken in the standard HRS calibration plan.  This usually consisted of a few bias frames, 
flatfields and ThAr arc observations taken before and after the nights observing, however for one night no flats were observed and a flatfield from a night close to the observing 
night was used in the reduction.  Also the number of flatfields were increased to 10 for the last six months or so of data, which significantly aids in fringe removal.  Indeed, the 
fringing in the red chip of HRS can reach 10-20\% and hence well exposed, high S/N Halogen flatfields are used to correct these.  

The \sc starlink \rm package \sc echomop \rm (\citealp{mills96}) was 
used to extract the echelle data.  Firstly, all 14 orders were located and traced using the master flatfield, and any stray pixels were clipped from this trace.  The dekker limits 
were determined using the master flatfield to ensure the entire object order and the adjoining sky order are included in the reduction.  The master flatfield was then 
normalized to determine the balance factors and these were multiplied into all other observations.  The sky light was removed by utilising the sky fiber and subtracting out any sky 
lines that appeared in the object spectra.  The orders were then extracted using an optimal extraction routine (\citealp{horne86}) and wavelength calibrated using the ThAr arc images.  

A small portion of spectra for the M4.5V star G~121-028, centered on the sodium doublet at 8182-8194\AA, is shown in Fig.~\ref{spectra} (solid line).  The 
dashed spectrum is the telluric standard HD1839, which is a fast rotating B star, therefore all the strong features are due to telluric contamination.  The 
sodium doublet is heavily blended with a number of telluric features, which without good sky subtraction, could serve to contaminate any 
measurement of the line properties.  This is one of the reasons that we use a spectral deconvolution routine to help remove the telluric features in our final analysis.  These final spectra 
have a limiting resolution of between $\sim$0.20-0.35\AA\ ($\sim$7-13~km~s$^{-1}$), which was determined individually for each star by deconvolving the telluric lines in each image.

\section{$v$~sin~$i$ Determination}

\begin{figure}
\vspace{2.5cm}
\hspace{-4.0cm}
\includegraphics{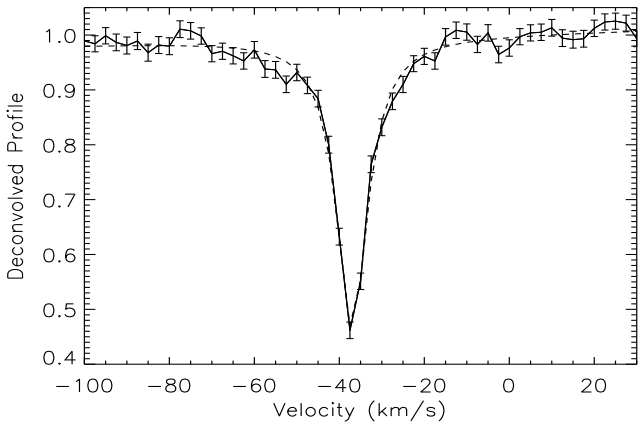}
\vspace{0.5cm}
\caption{The deconvolved line profile for the star G~121-028 in velocity space (vacuum velocities).  The best fit to this profile is marked by the dashed line. }
\label{decon_fit}
\end{figure}

All $v$~sin~$i$s in this work are determined by deriving an optimal line profile for each star, and then convolving a non-rotating template (LHS1950 is our 
template)  with rotational profiles of various velocities to find the best match.  A deconvolution 
method (\citealp{donati97}; \citealp{barnes98}; \citealp{cameron02}) is used to determine the optimal line profile.  A continuum fit is performed using 
the \sc continuum \rm routine in \sc iraf \rm to normalize the spectra and to provide inverse variance weights for the deconvolution procedure.  This least squares 
deconvolution uses a line list which best matches the spectral type of each star. The wavelength positions and line depths used were taken 
from VALD (Vienna Atomic Line Database; \citealp{kupka99}).  Any deficiencies in the line list data for cool M stars, such as missing opacities, should not affect the final $v$~sin~$i$ values since the same list is used on the non-rotating template star LHS1950.  The use of fewer model lines for deconvolution than actually observed will simply give rise to profiles with lower 
S/N than a more complete model would afford.  The line strengths of the final profiles may also vary from star to star but can be corrected for by scaling the non-rotating template to 
the line strength of each stellar deconvolved profile.  The method of least 
squares deconvolution is used to determine the mean line profile which when convolved with the line depth pattern gives the optimal match to the observed spectrum.  

\begin{figure}
\vspace{4.0cm}
\hspace{-4.0cm}
\includegraphics{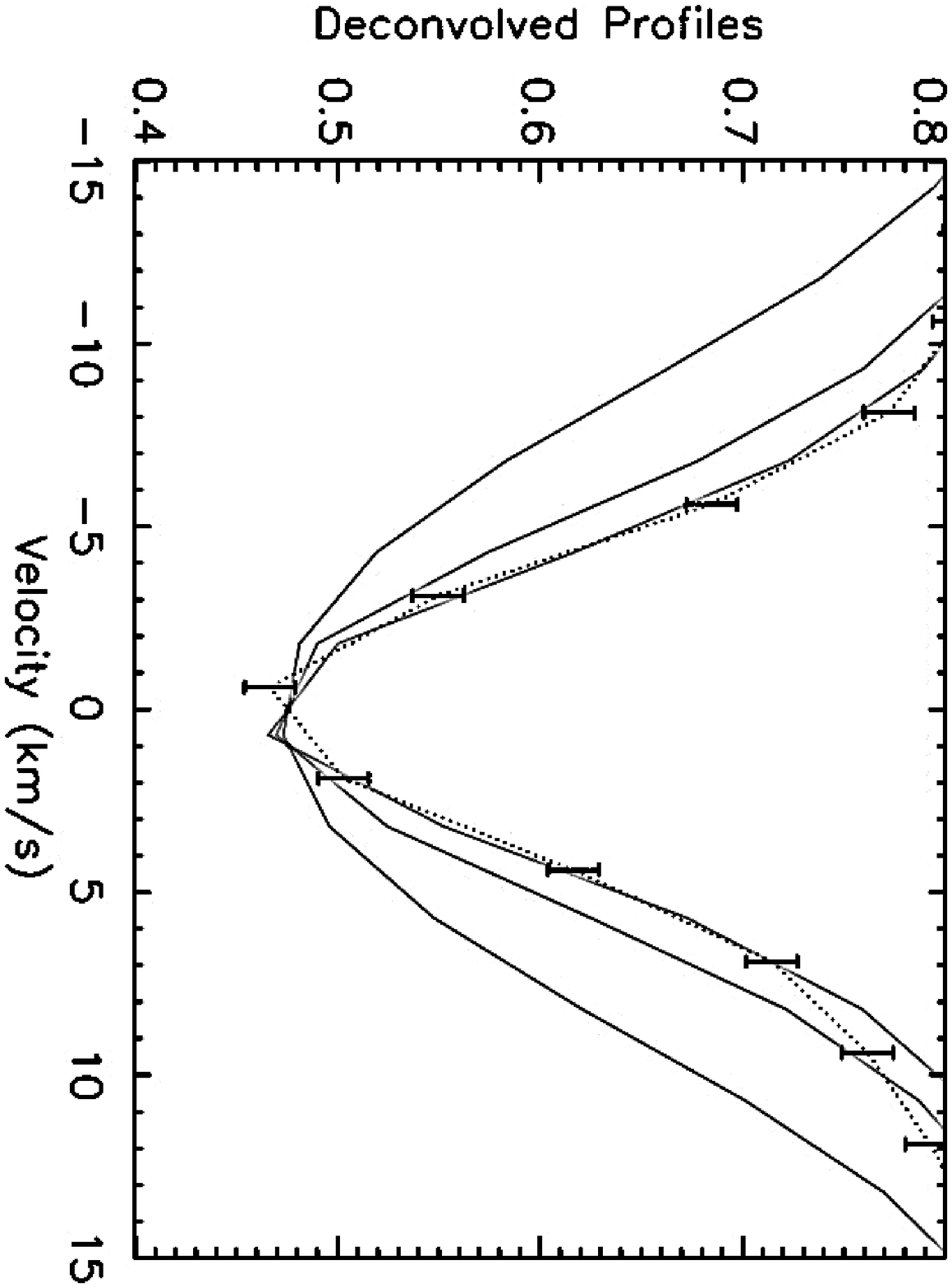}
\vspace{0.5cm}
\caption{A zoomed in region of deconvolved profile of G~121-028 (dotted curve), along with three broadened template profiles (solid curves).  From inner to outer the template has 
been broadened by velocities of 3.5, 5.0 and 10.0~km~s$^{-1}$ respectively.}
\label{decon_conv}
\end{figure}

Fig.~\ref{decon_fit} represents the final deconvolved line profile for G~121-028 (solid line) with velocity on the $x$-axis and normalized flux on the $y$-axis.  
G~121-028 is one of the slower rotating stars in our sample ($v$~sin~$i$=3.8$\pm$0.7~km~s$^{-1}$) as the narrow line profile shows.  To determine the $v$~sin~$i$ 
values we perform a Levenberg-Marquardt least squares minimisation using MPFIT in IDL (\citealp{markwardt09}) to find the best fit for each observed line profile.  The fit is represented 
by the dashed line in the figure and for this example a reduced $\chi$$^2$ ($\chi$$_{\nu}$$^2$) of 1.2 is found.  This best fit is used to determine both the profile centroid and scaling factor.  
The centroid allows each star to be shifted to the rest frame for use in the template comparison, whilst the scaling factor allows us to scale our non-rotating template to match 
the profile strength of each star.  LHS1950 was found to be our narrowest profile (determined from the full width at half maximum (FWHM) of the best fit to each profile) and 
hence below the resolution limit of the instrument.  This template is then broadened in steps of 0.5~km~s$^{-1}$, between velocities of 0.5-50~km~s$^{-1}$, using a grid of rotational 
profiles with a limb darkening coefficient of 0.6.  An example is shown in Fig.~\ref{decon_conv}.  A zoomed in region of the profile of G~121-028 is represented by the dotted curve, with template profiles marked by the 
solid curves.  From inner to outer (narrowest to broadest) the solid curves have been broadened by rotational profiles of 3.5,~5.0~and~10.0~km~s$^{-1}$ respectively.  This highlights the 
good fit between the inner profile compared to the outer profiles within the uncertainties and highlights the robust nature of the fitting technique.  All rotational profiles are generated 
following the description in \citet{gray92}, and a $\chi$$^2$ value is determined at each $v$~sin~$i$ step.  

\begin{figure}
\vspace{5.0cm}
\hspace{-4.0cm}
\includegraphics{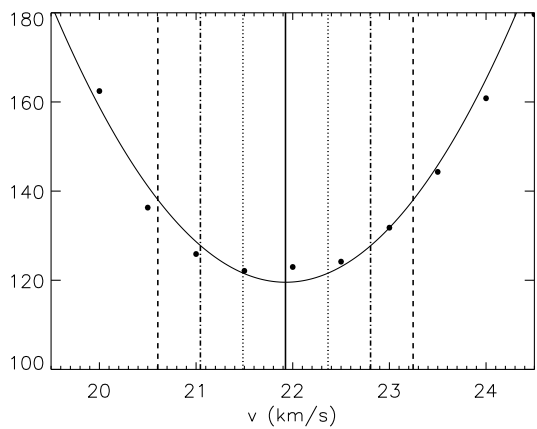}
\vspace{0.5cm}
\caption{A subsection of the best fit $\chi$$^2$ curve to the data (filled circles) for the star G~180-011.  The solid curve marks the best fit, with the minimum of this function marked by 
the vertical solid line.  The observed rotational velocity is 21.9~km~s$^{-1}$ for this star.  The dotted, dot-dashed and dashed lines mark the determined 1$\sigma$, 2$\sigma$ 
and 3$\sigma$ limits respectively.}
\label{vsini_chi}
\end{figure}

Fig.~\ref{vsini_chi} shows our $\chi$$^{2}$ fit to the data points in the analysis of the star G~180-011.  Each filled circle represents one of the 0.5~km~s$^{-1}$ steps used to broaden the 
template to fit the profile of G~180-011.  The solid curve represents the best quadratic fit to the data, and following the procedure in \citet{jenkins08}, 
the minimum of this curve is the measured $v$~sin~$i$, highlighted here by the vertical solid line.  The vertical dotted, dot-dashed and dashed lines represent the 1$\sigma$, 2$\sigma$ and 3$\sigma$ internal uncertainties respectively.  The 1$\sigma$ uncertainty value was determined by 
measuring the width of the curve at a 1$\sigma$ step (i.e. $\chi$$^{2}_{1\sigma}$ = $\chi$$^{2}_{min}$ + 1), after the curve has been broadened by 
a factor determined from the difference between the minimum of the $\chi$$_{\nu}$$^{2}$ and 1 (where $\chi$$_{\nu}$$^{2}$ is the best fit).  Due to the nature of the dataset this factor has a 
median of only 2.1 and standard deviation of 1.5, indicating the goodness of the methodology.  We employ this artificial broadening to 
the $\chi$$^{2}$ curve to help alleviate uncertainties that are difficult to address in the analysis procedure e.g. macroturbulence variations, spectral type differences, etc.  For G~180-011 
shown in the figure the formal uncertainty is found to be $\pm$0.4~km~s$^{-1}$.  We use this procedure to determine all 1$\sigma$ uncertainties quoted in Table~\ref{tab:vsini}.  

The 1$\sigma$ uncertainties mentioned show we have good internal precision and since we have a few stars where multiple observations were taken we can test this in a brute force manner by 
looking at the values for these multiple measurements individually.  When we do this we find that the $v$~sin~$i$s tend to agree to within the uncertainties determined by the $\chi$$^{2}$ fitting procedure on each individual measurement.  
This gives us confidence in the analysis procedure.  We note that for the star GJ1253 the $v$~sin~$i$ values agree to within 1$\sigma$, which for this star was $\sim$1~km~s$^{-1}$, 
and as these were measured over two observations a month apart, they are probably the best indicator of the overall random uncertainties in the analysis procedure.  Therefore, all 
velocities for objects $\le$10~km~s$^{-1}$ are accurate to $\pm$1~km~s$^{-1}$.  For the objects rotating much faster than this, particularly above 20~km~s$^{-1}$, it is difficult to 
fit their profiles to better than $\sim$$\pm$10-20\% at the adopted resolution and S/N and therefore this should be taken as the accuracy for such fast rotators.

\subsection{Instrumental Profile}

Although we estimate our resolution limit to be around 2.5~km~s$^{-1}$, we note that this is not the case for all of our sample.  Since we use a fixed template spectrum that 
was measured on our first night and not on every night of our observations, there will be effects due to the changes of the instrumental response.  To monitor this 
we used the telluric lines in each of our spectra as a proxy for the instrumental profile (IP).  We assume that the intrinsic width of weak telluric lines 
are smaller than the instrumental resolution.  We then performed the same fitting routine to these deconvolved profiles which returned an instrumental FWHM 
for each stellar spectrum.  We 
decided not to use the arc lines for the IP measurements since the calibrations are taken through a different optical path than the science images and therefore 
would not accurately model the intrinsic instrumental width (see \citealp{tull98} for HRS design).  This procedure allowed us to determine that there was an increase in the instrumental 
width of almost a factor two between 2007 August 9$^{\rm{th}}$ and 2007 October 17$^{\rm{th}}$ due to poor thermal control in the HET spectrograph that lead to PSF instability through 
focus drift.  The stars observed between these dates have a typical resolution limit of around
4-5~km~s$^{-1}$, therefore we employ a correction to each star to correct for the effects of the broadening of the IP.  

\begin{figure}
\vspace{5.0cm}
\hspace{-4.0cm}
\includegraphics{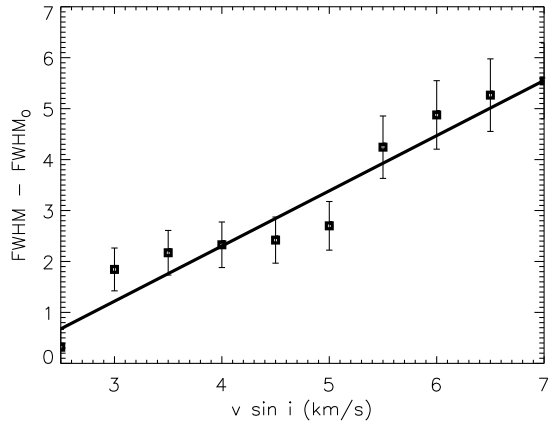}
\vspace{1.3cm}
\caption{The change in the FWHM as a function of $v$~sin~$i$.  The squares show the average FWHM minus the initial FWHM (FWHM$_{0}$) after the 
profile has been broadened in steps of 0.5~km~s$^{-1}$.  The solid curve is the best straight line fit to the data.}
\label{fwhm_fit}
\end{figure}

Fig.~\ref{fwhm_fit} shows the average change in the FWHM of a selection six stars with the highest S/N ratios as a function of $v$~sin~$i$.  Each of the data points mark the average of the 
FWHM difference (FWHM-FWHM$_{0}$) for the six stars, along with the associated scatter of the values, after broadening by rotation profiles in steps of 0.5~km~s$^{-1}$, where FWHM$_{0}$ 
is the initial FWHM before any broadening has been applied.  
The solid line shows the best straight line fit to the data that we employed to correct for the IP, and is described by 1.08$x$-2.03, with an RMS scatter of only 0.46~km~s$^{-1}$.  

To employ these corrections 
we determine the difference between the template telluric FWHM and that of each star.  We then use the fit to determine which $v$~sin~$i$ profile we must broaden the template 
profile by initially to correct for the changing IP width.  Due to our resolution this only becomes significant on the data mentioned above where the IP almost doubled in 
width, leaving such stars with a resolution of 4.5~km~s$^{-1}$.  These fits can also be used to determine the actual $v$~sin~$i$ from the change in the FWHM compared to the template, 
at least in the low $v$~sin~$i$ regime, and a test of this reveals agreement between both methods.  The fits could be extended into the high $v$~sin~$i$ regime in order to use this 
method to determine the rotation velocities for rapidly rotating stars also.  All $v$~sin~$i$s and their associated uncertainties are shown in column~11 of Table~\ref{tab:vsini}, with the 
telluric FWHMs shown in column~12 for reference.  

\begin{figure}
\vspace{5.0cm}
\hspace{-4.0cm}
\includegraphics{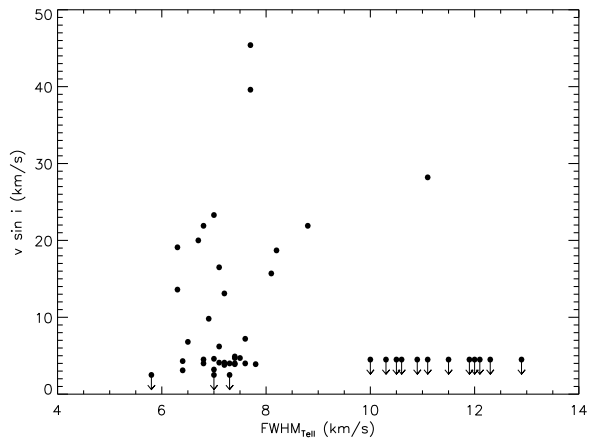}
\vspace{1.3cm}
\caption{The distribution of $v$~sin~$i$ as a function of the change in width of the instrumental profile.  The width is parameterised by the FWHM of the telluric lines in km~s$^{-1}$.  
No correlations are evident in the spread meaning the correction applied to the template spectrum appears robust.  The symbols with downward pointing arrows indicate upper 
limits.}
\label{vsini_tell}
\end{figure}

Fig.~\ref{vsini_tell} shows the overall spread in the measured $v$~sin~$i$ values as a function of the changing IP of the instrument.  As mentioned above the change in the IP 
is monitored by measuring the FWHM of the telluric lines in each spectrum (FWHM$_{\rm{Tell}}$) and from the plot there appears no significant correlation between measured rotation 
values and the width of the IP.  The linear correlation coefficient (r) for the entire sample is -0.12 and is still only 0.36 when removing all stars with measured upper limits, which are 
represented in the figure by the downward pointing arrows.  This test highlights the lack of any significant correlation between the measured $v$~sin~$i$s and the changing width of the IP.  In fact, only one of the stars measured during the period of increased IP 
values is found to be above the detection threshold of our method, and as will be seen later, this correlates with the presence of H$\alpha$ in emission in this star.

None of these objects have previously determined $v$~sin~$i$s with which to compare our 
values for an accuracy check.  However, \citet{rockenfeller06} determined the $I$-band variability for the star LHS2930 and found the period to be 
13$\pm$2hrs.  They also quote the 
radius of the star to be 0.33R$_{\odot}$, from which we can estimate the rotation velocity assuming the variability is induced by the rotation of surface features such as star spots.  
Estimating this velocity in pure spherical geometry returns a value for the equatorial rotation velocity of $\sim$30$^{+6}_{-3}$~km~s$^{-1}$.  We measure a velocity of 18.7$\pm$1.5~km~s$^{-1}$, which is significantly lower than the photometrically derived value.  We note 
that the deconvolved line profile for this star is flagged as low S/N.  However, given these values this would indicate the star is inclined to our line of sight by 
around 51$^{+10}_{-7}$$^{\rm{o}}$ ($i$$\sim$39$^{\rm{o}}$).  \citet{marcy96} show that for random stellar alignments, 68\% of 
stars will have inclinations to our line of sight of below 47$^{\rm{o}}$, with 95\% having inclinations above $\sim$20$^{\rm{o}}$, placing LHS2930 in between this 1 and 2$\sigma$ result and 
in agreement with the 1$\sigma$ value within the estimated uncertainty.

\subsection{H$\alpha$ Absorption/Emission}

In addition to selecting against stars with a large $v$~sin~$i$, one might also like to monitor each star's magnetic activity through the absorption or emission in the H$\alpha$ line.  The 
correlation between H$\alpha$ emission, age, activity and rotation velocity has been well studied (e.g. \citealp{mohanty03}; \citealp{reiners08} and refs therein), however the magnetic field 
structure of M dwarfs, particularly slowly rotating M dwarfs, could be complex, with magnetically active regions rotating in and out of view and the presence of strong flaring events.  Also there 
is the possibility that a small percentage of our M dwarfs are being viewed close to pole-on and instead of measuring their true $v$~sin~$i$, which could be very high, we measure a smaller 
value and place these in the good planet search target bracket.  Therefore, we searched for any emission, and also significant absorption, in the H$\alpha$ line ($\lambda$$\sim$6562\AA) in 
each of our target stars.  24 (43\%) stars were found to show H$\alpha$ emission, which includes the double profiles found for the active binary stars, and since such stars could be 
young and highly active, these may not be ideal stars to include in a precision radial-velocity planet search.  Of these only six have rotation velocities below 10~km~s$^{-1}$ (12\% 
of the total).   However, we do 
caution that we might be viewing active regions on these stars, or flaring outbursts, since numerous M stars across the spectral domain have been found to exhibit such phenomena 
(e.g. see \citealp{tinney98}; \citealp{martin01}; \citealp{osten05}).   Further measurements may reveal these to vanish and therefore more epochs may be 
required to test if such stars have continuous H$\alpha$ emission and are probably highly inclined young and active stars, or are simply going through a flaring event.  
\citet{reiners09} has shown that flares only create significant noise at velocity precisions below 10~m~s$^{-1}$ for moderate events and at the level of a few hundreds of m~s$^{-1}$ for 
giant flares.  However, giant flares are also heavily correlated with H$\alpha$ variation and so can be easily removed from a radial-velocity campaign, albeit at the cost of precious observing 
time.  Indeed, we may have viewed such a scenario in one of our stars GJ1253.

\begin{figure}
\vspace{5.0cm}
\hspace{-4.0cm}
\includegraphics{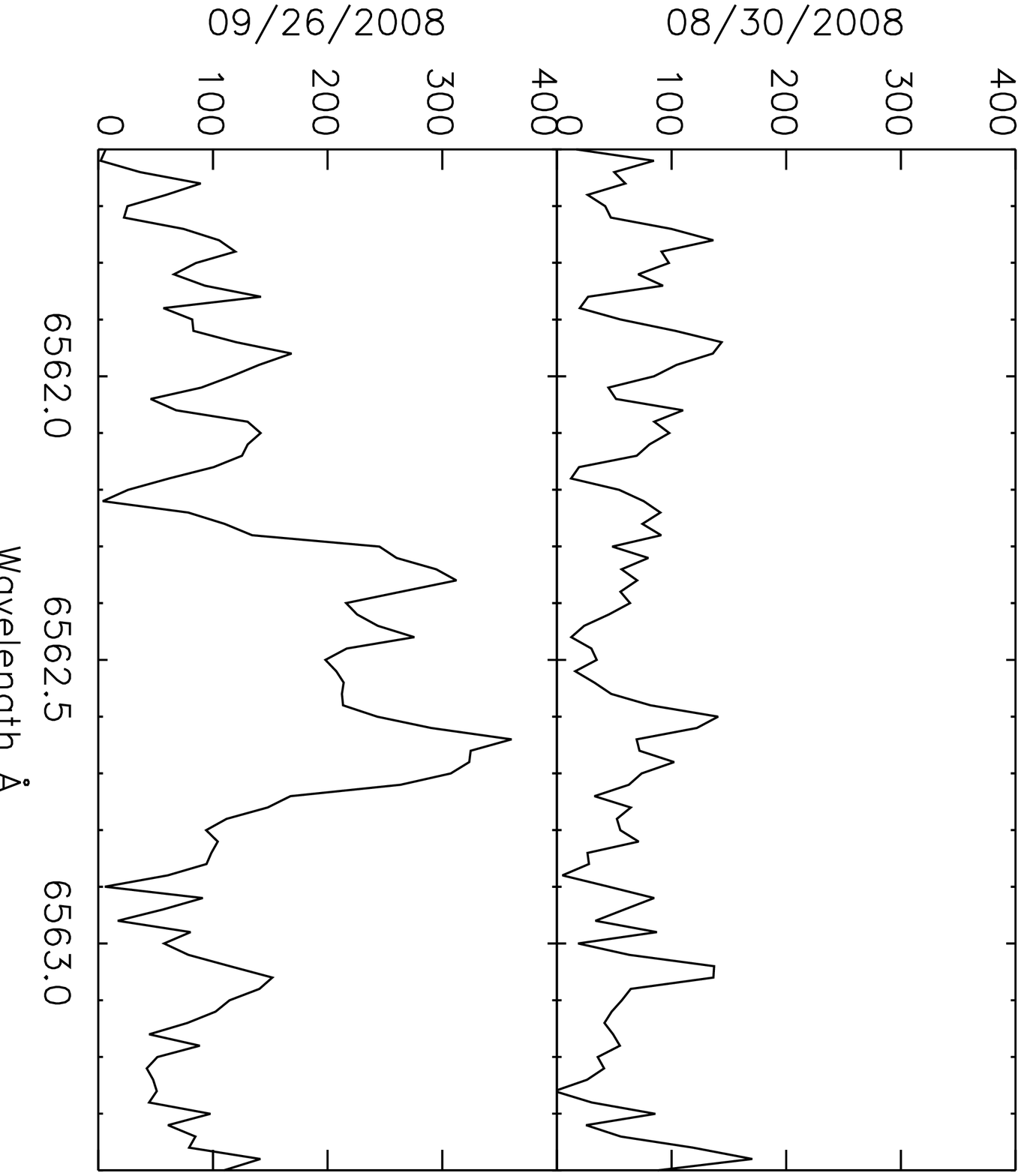}
\vspace{1.3cm}
\caption{Region around the H$\alpha$ line profile for the star GJ1253 at two separate epochs.  The top plot shows the first measurement made on 08/30/2008 and no H$\alpha$ emission was found.  However, 
the lower plot shows the second measurement made for this star one month later (09/26/2008) and clear emission is present.}
\label{flare}
\end{figure}

Fig.~\ref{flare} shows two epochs of spectroscopic measurements of the region around the H$\alpha$ line for the star GJ1253.  The upper panel is the first observation, 
made on 08/30/2008, whereas the lower panel shows the same region only one month later (09/26/2008).  In the first observation there is no indication of any H$\alpha$ 
emission present, however only 27 days later when the second measurement was obtained, there is clear evidence for the H$\alpha$ line in emission.  Such a short period 
evolution of the line, coupled with the relatively low $v$~sin~$i$ of the star ($\le$~4.5~km~s$^{-1}$) may indicate that a small magnetic region is rotating in and out of 
view.  If confirmed continuous monitoring of such H$\alpha$ line changes could lead to the true rotation period of the star, 
removing the sin~$i$ degeneracy and giving both the true velocity and, for any discovered companions, their true masses without the need 
for continuous photometric monitoring.  Note that \citet{byrne96} found variable H$\alpha$ emission in the active M star HK Aquarii which they attribute to prominence 
like clouds above the stellar surface that have characteristic timescales less than half the photometrically determined rotation period.  It may be the case that such 
variability of H$\alpha$ flux is mainly generated through flares or a changing magnetic field structure.  Indeed, H$\alpha$ variability for M stars is significant across the spectral 
domain, even on relatively short time scales ($<$60~minutes).  \citet{lee09} find that $\sim$80$\%$ of their mid-to-late M star sample exhibited statistically significant 
H$\alpha$ variability.  Therefore, if any correlation does exhist between H$\alpha$ variability and rotation velocity the characteristics must be sufficiently large such that 
typical short period small scale variations do not mask it out.

Finally, we also searched for significant H$\alpha$ absorption and found eight stars (14\%) that show significant absorption.  All of these have $v$~sin~$i$s that would meet our criteria for planet search 
selection, 
as would be expected from a rotation-activity connection.  Only, 2 out of the 8 objects that exhibit significant H$\alpha$ absorption are below our resolution limit, with a further 3 in 
agreement with the adopted resolution limit to within the estimated total uncertainty of the analysis technique.  All stars with no H$\alpha$ emission would 
meet our criteria for planet search selection, with the largest 
rotation rate of these coming from the M5 dwarf GJ1182 ($v$~sin~$i$=6.8$\pm$0.5~km~s$^{-1}$).  Column~13 of Table~\ref{tab:vsini} shows the H$\alpha$ emission flag, where either a detection 
was made or not, and stars with no significant emission which have asterisks show evidence for H$\alpha$ absorption.

\subsection{Spectroscopic Binaries}

\begin{table*}
\small
\center
\caption{M dwarf binary candidates}
\label{tab:binary}
\begin{tabular}{cccccccccc}
\hline
\multicolumn{1}{c}{Star}& \multicolumn{1}{c}{$V$}& \multicolumn{1}{c}{$J$}& \multicolumn{1}{c}{$H$}& 
\multicolumn{1}{c}{$K_{\rm{s}}$} & \multicolumn{1}{c}{Spec Type}& 
\multicolumn{1}{c}{$\Delta$$v$~(km~s$^{-1}$)} & \multicolumn{1}{c}{$v$~sin~$i$$_{\rm{pri}}$~(km~s$^{-1}$)} & \multicolumn{1}{c}{$v$~sin~$i$$_{\rm{sec}}$~(km~s$^{-1}$)} & \multicolumn{1}{c}{H$\alpha$ Emission} \\ \hline
GJ1080  & 12.81 & 8.98 & 8.50 & 8.22 & M3.0 & 28.44$\pm$0.28 & $\le$3.0 & $\le$3.0 & No  \\
GJ3129  & 14.27 & 9.65&  9.06 & 8.80 & M4.5 & 89.36$\pm$0.78 & 5.2$\pm$0.1 & 6.6$\pm$0.3 & Yes - Double  \\
Gl802   & 14.67 & 9.56&  9.06 & 8.75 & M5.0 &150.03$\pm$1.19 & 6.4$\pm$0.4 & $\le$4.5 & Yes - Double   \\
LHS3080 & 14.28 & 9.67 & 9.11 & 8.82 & M4.5 & 18.35$\pm$0.32 & $\le$3.0 & 3.8$\pm$0.1  & Yes - Single \\
\hline
\end{tabular}
\medskip
\end{table*}

\begin{figure}
\vspace{3.5cm}
\hspace{-4.0cm}
\includegraphics{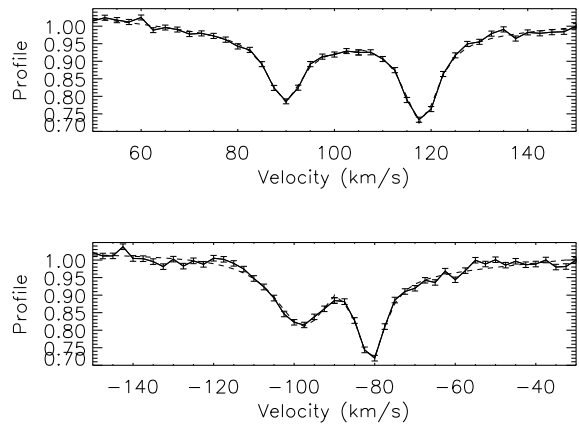}
\vspace{1.3cm}
\caption{Example blended deconvolved profiles for two of the probable binary systems in this sample.  The upper plot represents the star GJ1080 and has two clearly defined deep profiles, whereas 
the lower plot, which represents the LHS3080 system, has a secondary profile also, possibly from a weak signal contribution from the secondary.  The dashed 
curves represent the best fit double profiles to the data.  The two other potential binaries have profiles with much larger separations.}
\label{binary}
\end{figure}

There is the possibility that stars with miss shapen profiles are spectroscopic binaries of similar spectral type or stars with powerful magnetic fields.  Fig.~\ref{binary} shows blended 
profiles found for both GJ1080 (upper panel) and LHS3080 (lower panel).  It is clear that for both of these systems there is a primary deconvolved profile and a weaker 
secondary.  To test if these are real rather than deconvolution artifacts, we split the spectra into two wavelength ranges, one spectrum covering the blue orders 
and one covering the red orders. By deconvolving both the red and blue spectra independently, we were able to confirm that these stars still exhibit double line profiles, lending 
weight to fact that the double profiles are due to real phenomena.  The spectra for GJ1080 shows double-lined profiles for atomic and molecular 
lines.  The double-lined profiles for LHS3080 were difficult to confirm by eye in the spectrum.  Along with both of these, there are another two binary systems, those of GJ3129 and Gl802.  
Both of these stars have double profiles, but with much wider separations, even though they exhibit noisier spectra.  These are more clear cases than the previous two since they also have 
double H$\alpha$ emission profiles.

Gl802 was known to have a low-mass companion with an orbital period of 3.14$\pm$0.03~yrs (\citealp{pravdo05}), which was directly imaged by \citet{lloyd06}.  In addition, \citet{ireland08}  
has recently found this star to be part of a triple system, with a short period ($P$$\sim$19~hrs) spectroscopic companion.  Our final deconvolution for this star is very noisy, in part due 
to the blended light from three different components with different spectral profiles.  We do note however that there are two strong profiles in the final deconvolution (properties listed 
in Table~\ref{tab:binary}) and potentially a further two weaker profiles.  The weak profiles are only borderline significant due to the associated noise, but one might explain the short 
period binary found by Ireland et al. and the other, which is widely separated from the profile of Gl802A, could be an additional longer period companion.

\citet{mccarthy04} surveyed both GJ3129 and LHS3080 as part of their AO campaign to detect 
substellar companions to a host of nearby young M dwarfs.  They found no viable candidate to these stars between around 5$''$ to 15$''$ arcseconds.  
The lack of any viable detection around these, apparently young, stars is probably explained by the close separation of the objects, since the detectability of spectroscopic binaries are 
skewed towards short period companions.  Indeed, the LHS3080 profiles are
blended with each other and since the secondary profile is significantly
weaker than the primary, it is probably a much later and fainter companion object.  The properties of these spectroscopic binaries are listed in Table~\ref{tab:binary}, which 
include the Hipparcos $V$, 2MASS $J$, $H$ and $K_{\rm{s}}$ apparent 
magnitudes, their spectral types, the velocity separation of the double peaks along with their combined uncertainties, the $v$~sin~$i$ estimates for both the primary star 
($v$~sin~$i$$_{\rm{pri}}$) and the secondary ($v$~sin~$i$$_{\rm{sec}}$) and the H$\alpha$ flag.  Note the small uncertainties are an artifact of the blended fitting procedure and 
do not necessarily mean these values are more precise than the single profiles.

\section{Temperature, Mass and Metallicity}

When performing radial-velocity searches for exoplanets other characteristics of the star determine radial-velocity limits e.g. the stellar mass.  The mass of the primary 
determines the lower limit to the planetary mass for a given radial-velocity amplitude and therefore knowledge of the mass of any planet search target star is essential.  We have 
determined the stellar mass for the bulk of our sample by utilising relations between M star mass and their absolute magnitudes.  To generate accurate absolute magnitudes we searched 
the Yale Trigonometric Parallax project (\citealp{altena95}), the RECONS list (\citealp{henry06}) and any parallaxes in the Gliese Catalogue of Nearby Stars to obtain accurate parallax 
measurements for all our candidates.  The parallaxes, along with their associated 
uncertainties, are shown in column~8 of Table~\ref{tab:vsini} and these were used to determine the 
absolute magnitudes by converting them to distance and measuring the distance modulus.  The photometry was acquired using both the 
Simbad$\footnote{Simbad website: http://simbad.u-strasbg.fr/simbad/}$ and Vizier$\footnote{Vizier website: http://webviz.u-strasbg.fr/viz-bin/VizieR}$ astronomical databases.  The near 
infrared photometry was taken from the 2MASS catalogue (\citealp{strutskie06}) and the $K_{\rm{s}}$ magnitudes were converted to $K$ using the calibrations in \citet{carpenter01}.  The 
absolute $V$ and $K$ magnitudes were then input into the empirical mass relations from 
\citet{delfosse00}, and two measurements of the mass for each star were determined.  Both these measurements were averaged to get the final combined mass estimates for all objects 
with known parallax and their uncertainties were taken as the standard deviation of the two measurements.  The uncertainties are typically $\pm$10-15\%, with a few of the closest 
stars having uncertainties down to the $\pm$1\% level.  Both the masses and their uncertainties are shown in column~9 of Table~\ref{tab:vsini}.  

Along with the mass, both the 
effective temperatures and metallicities ([Fe/H]) were estimated for the sample.  The effective temperatures were determined using the $V-K_{\rm{s}}$ relation taken from \citet{casagrande08} and 
has a typical internal uncertainty of $\pm$17K, however the uncertainty on the overall accuracy of the technique will probably be substantially larger than this.  The values with question marks 
next to them flag highly suspect effective temperatures.  The [Fe/H] abundances 
were firstly determined photometrically using the colour magnitude relation in \citet{bonfils05}.  They quote the typical uncertainty for this method as $\pm$0.2~dex which we assign to 
all our metallicities 
shown in column~10 of Table~\ref{tab:vsini}.  Casagrande et al. also provide metallicities using their method and these agree with the Bonfils et al. values within the quoted 
uncertainties.  Note that recently \citet{johnson09} claim that the Bonfils et al. relation may underestimate the metallicities of M dwarfs by as much as 0.3~dex.  Finally, 
Table~\ref{tab:vsini} also shows each star's $V, J, H$ and $K_{\rm{s}}$ photometry.

\section{Results: $v$ sin $i$ Distributions}

\begin{figure}
\vspace{4.5cm}
\hspace{-4.0cm}
\includegraphics{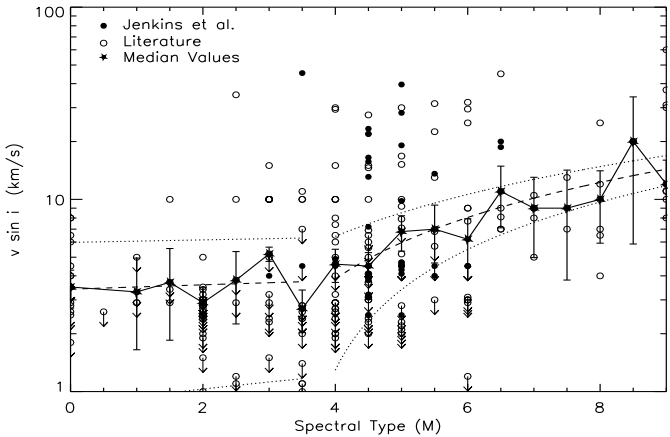}
\vspace{0.5cm}
\caption{The distribution of $v$~sin~$i$s against measured spectral type plotted in log step.  The values from this data are shown by filled circles, with the literature values shown by open circles.  
The trend of increasing rotation velocity with decreasing temperature is seen here and is highlighted by the solid line linking the filled stars.  These stars mark the
median values for all 
data in each spectral bin.  The uncertainties plotted represent Poisson statistics and the downward pointing arrows mark upper limits.  Represented by the dashed lines are the best 
straight line fits to the medians for the early M dwarfs and mid-to-late M dwarfs, along with their standard $\pm$1$\sigma$ uncertainties plotted by the dotted lines.  The 
early Ms are found to have a much flatter distribution than the mid-to-late Ms, which are tending towards a more rising trend.
}
\label{vsini_plot}
\end{figure}

Fig.~\ref{vsini_plot} shows the distribution of early-to-late M star rotation rates against their spectral type.  The filled circles represent the stars in this work and the open 
circles represent literature $v$~sin~$i$s (from \citealp{stauffer86}; \citealp{marcy92a}; \citealp{delfosse98}; \citealp{gizis02}; \citealp{mohanty03}; \citealp{bailer-jones04}; \citealp{fuhrmeister04};
\citealp{jones05}; \citealp{reiners07b}; \citealp{west09}).  All literature $v$~sin~$i$s are shown in Table~\ref{tab:lit_vsini} along with any measured $V$ magnitudes, 2MASS $J$, $H$ and $K_{\rm{s}}$ 
photometry, spectral types, effective temperatures, parallaxes, masses and [Fe/H] abundances, all determined using the same methods outlined above.  

There is a large spread in $v$~sin~$i$ across the mid M star regime (M4.0-M6.5) compared with early M stars ($<$M4).  The 
mid M objects range from almost as high as over 50~km~s$^{-1}$ and down as low as
essentially zero km~s$^{-1}$.  The median dispersion of the $v$~sin~$i$ distributions in the spectral type bins between M0-M3
is 3.70$\pm$0.79~km~s$^{-1}$, compared to the median dispersion of 9.00$\pm$8.31~km~s$^{-1}$ for bins at M4-M9.5.  However, if we allow for a spectral typing uncertainty of $\pm$0.5 
sub-types and exclude both the M0 and M9.5 spectral bins, since they could be contaminated by K and L dwarfs, we find the dispersions are 3.70$\pm$0.87~km~s$^{-1}$ and 9.00$\pm$4.37~km~s$^{-1}$ 
respectively.  A K-S test is run to compare these distributions and this returns a D-statistic of 0.539, giving an extremely low probability of only 1.396x10$^{-8}$\% that the rotation rates 
for stars between M0.5-M3 are drawn from the same parent population as those between M4-M9.  Also, a large fraction of stars with spectral types below M6.5, and particularly below the 
convective boundary region, only have measured upper limits.  This will bias this result towards a non-correlation and so we can expect that with further detections at low velocities for 
these stars, this correlation will become more pronounced.  

This step from low to high dispersions between M3 to M4 spectral types is thought to be due to the 
increased spin-down timescale towards decreasing mass (\citealp{delfosse98}; \citealp{mohanty03}) and that the spin-down timescale is a significant fraction of the age of the young disk.  
Indeed, \citet{donati08} and \citet{morin08} have shown that the magnetic field topologies of early M's (M0-M3) and mid-M's (M4) are significantly 
different, with the early M's having mainly toroidal and non-axisymmetric poloidal fields, whereas the M4s mainly exhibit axisymmetric poloidal fields.  
This result indicates a change in the magnetic field properties at the classical boundary between partially radiative and fully convective envelopes.  If the spin down times 
are governed by the magnetic fields in this regime then a differing $v$~sin~$i$ distribution might be expected.  It might be the case that axisymmetric 
poloidal fields interact with the stellar wind more weakly than the toroidal, non-axisymmetric fields, driving a less efficient braking mechanism.  Interestingly, a probe of the spindown 
timescale and the mechanism driving it can be made if we take the Delfosse et al. 
and Mohanty \& Basri $v$~sin~$i$ and log(L$_{\rm{H}_{\alpha}}$/L$_{\rm{Bol}}$) values (Mohanty \& Basri show the activities in these two works to be in excellent agreement).  We can use 
these values to trace the rotational 
history for the fully convective, mid-M star regime (M4-M7).  When the stars are binned into their respective spectral types, the trend found between these two quantaties 
reveals a sharp saturation boundary (e.g. Fig.~9 in Mohanty \& Basri).  By using Eq$^{\rm{n}}$.~3.1 in \citet{west09b}, one can use the activity values to gain a statistical insight into the age distribution of these samples.  
In particular, the saturation boundary changes as a function of spectral type, increasing in velocity with increasing spectral type.  Table~\ref{tab:sat} shows the spectral type bins, along with 
the saturation boundary in velocity and activity lifetime ($l$) given in West et al.  Due to the spindown of stars by the wind braking mechanism we can use this table to say that from a statistical footing, the average age of M4 stars 
with $v$~sin~$i$s $\le$~8~km~s$^{-1}$ is older than 5~Gyrs, M5s with velocities $\le$~10~km~s$^{-1}$, and M6s with velocities $\le$~12~km~s$^{-1}$, are older than 7~Gyrs and M7s 
with velocities $\le$~13~km~s$^{-1}$ are older than 8~Gyrs. 

\begin{table}
\small
\center
\caption{M dwarf activity saturation limits}
\label{tab:sat}
\begin{tabular}{ccc}
\hline
\multicolumn{1}{c}{Spectral Type}& \multicolumn{1}{c}{$v$~sin~$i$ (km~s$^{-1}$)}& \multicolumn{1}{c}{$l$ (Gyr)}\\ \hline
M4  & 8     & 5 \\
M5  & 10   & 7 \\
M6  & 12   & 7 \\
M7  & 13   & 8 \\
\hline
\end{tabular}
\medskip
\end{table}

\begin{figure}
\vspace{5.5cm}
\hspace{-4.0cm}
\includegraphics{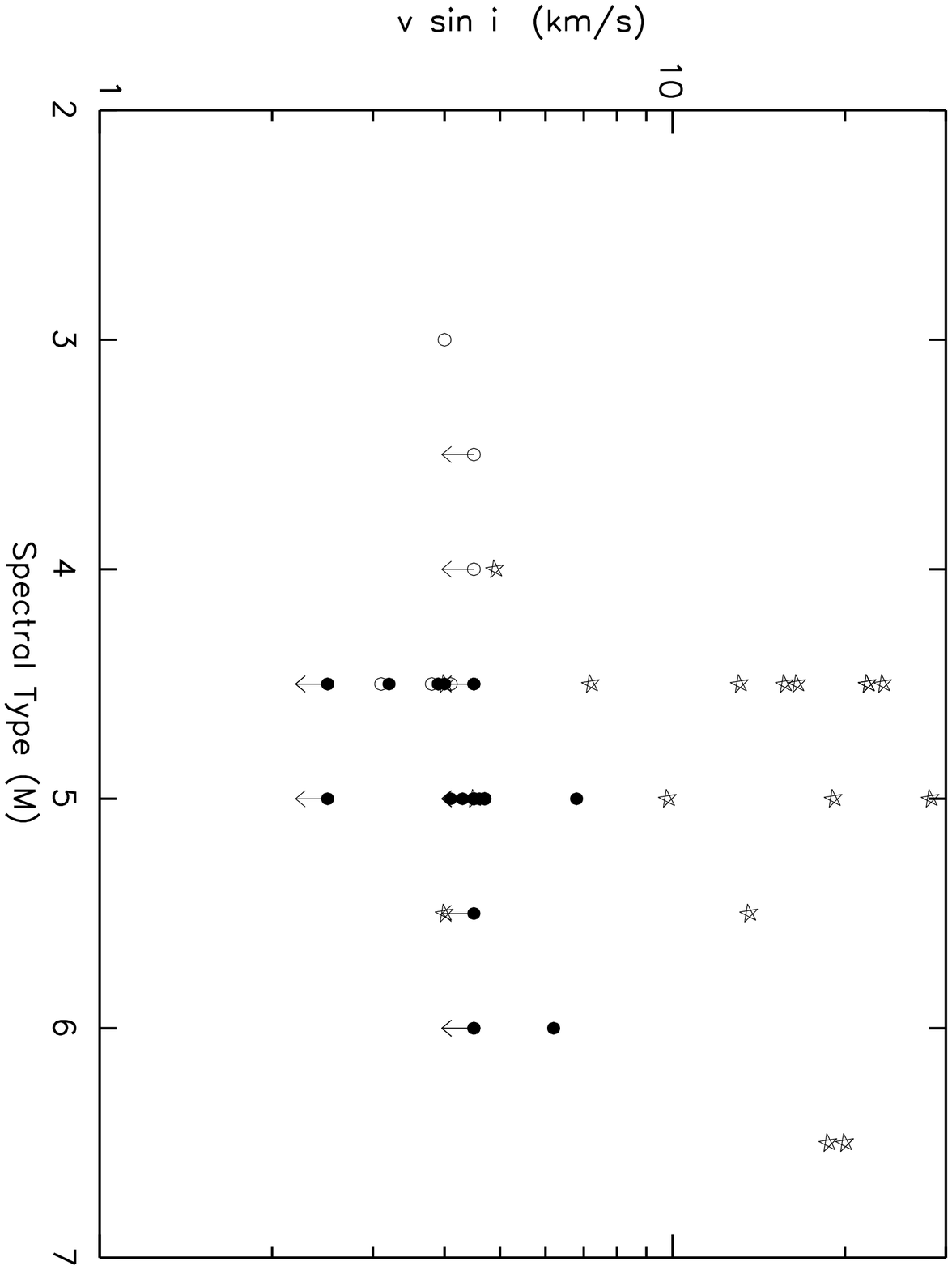}
\vspace{0.5cm}
\caption{The distribution of rotational velocities against spectral type for
stars in this study with a split based on each star's H$\alpha$ profile.  The filled circles represent stars with no significant H$\alpha$ emission or absorption.  
The open circles are stars with significant H$\alpha$ absorption and the stars represent objects with detected H$\alpha$ emission.  All stars 
with velocities greater than 7~km~s$^{-1}$ have detected H$\alpha$ emission.  Also no significant H$\alpha$ absorption was detected in stars later than M5.
}
\label{halpha_vsini}
\end{figure}

Fig.~\ref{halpha_vsini} shows the $v$~sin~$i$ values against spectral type
for all stars in this study with a split made by their H$\alpha$ status.  The filled circles represent all stars with no significant H$\alpha$ emission or absorption, 
the open circles represent stars with H$\alpha$ absorption and the open stars represent objects with H$\alpha$ emission.  The rotation-activity connection suggests a correlation between 
H$\alpha$ emission and the rotation velocity of cool M stars and this is clearly seen in this plot.  All stars with $v$~sin~$i$ $\ge$~7~km~s$^{-1}$ are found to exhibit significant H$\alpha$ emission.  Indeed, we have just shown that, on average, mid-to-late M stars with velocities $\ge$~8~km~s$^{-1}$ are still in their young, active phase of evolution.  No stars later than M5 were 
found to exhibit H$\alpha$ absorption, with five of these exhibiting no significant H$\alpha$ emission (55\%).  Such numbers correlate with previous results which suggest a high 
frequency of active M stars towards the latest spectral types 
(e.g. \citealp{fleming00}; \citealp{mohanty02}), even though we find a similar fraction of active M stars in all of our spectral bins.  However, in our final M6.5 bin both stars are found to exhibit 
H$\alpha$ emission which agrees with the increasing trend of activity towards values approaching 100\% at a spectral type of around M7.  This increase in stars with H$\alpha$ in 
emission towards later spectral types appears to be a product of the fraction of young disk stars in this regime and also the increase in activity lifetime with increasing spectral type.  
From Table~\ref{tab:sat} we see that both the activity lifetime and the time spent in a state of high rotation increases with increasing spectral type.  Such trends would naturally give rise 
to a larger fraction of stars with H$\alpha$ profiles in emission towards the latest M stars.

The filled stars in Fig.~\ref{vsini_plot} represent the median values in each spectral bin along with their associated Poisson errors.  The solid line connects the points and visually 
highlights the increasing trend towards later spectral types.  \citet{reid02b} suggest a flat distribution of rotation rates between stars of
M6-M9, whereas larger samples indicate this may not be the case (Mohanty et al.; \citealp{reiners08}).  The medians indicate a rising trend through the mid-to-late M dwarfs, flattening 
off towards the end.  Due to the change in distributions at the fully convective boundary we perform two straight line fits to the medians in the spectral ranges M0-M3.5 and M4-M9.  The 
best fit to the early M stars clearly shows a flat trend across the whole regime, whereas the fit to the late Ms show a rising trend, appearing to flatten towards the later M stars (note the 
curvature due to logarithmic plotting).  The fits to the early and late M star samples are described by $v$~sin~$i$=0.09($\pm$0.30)xSpT+3.41($\pm$0.67)~km~s$^{-1}$ and $v$~sin~$i$=2.10($\pm$0.53)xSpT-4.54($\pm$3.56)~km~s$^{-1}$, 
with standard uncertainties, represented by the dotted lines in the plot, of $\pm$0.89~km~s$^{-1}$ and $\pm$2.79~km~s$^{-1}$ respectively.  The total sample of less than 300 is 
still rather small, especially when binned.  Indeed there appears a dearth of objects between M6.5-M8.5 with rotation rates above 
$\sim$15~km~s$^{-1}$, which may indicate another population change above M6.5.  The evidence for this gap is weak at present due to the low number of stars in these spectral bins, 
therefore further observations and are needed and any biases studied to fully validate the existence of this feature. 

At temperatures below around 2800K (approximately M6-M7 type objects) dust formation and opacity are important in stellar/substellar atmospheres (\citealp{tsuji96}; 
\citealp{jones97}; \citealp{tinney98}; \citealp{chabrier00}; \citealp{baraffe02} and references therein).  \citet{berger08} have shown that late-M stars mark a transition in the 
properties of the magnetic field and its dissipation, along with high temperature plasma being generated in the outer atmosphere.  They go on to hypothesize that the stellar rotation 
may play a part in this process and indeed the difference shown here between the mid and late type M stars seems to add to this conclusion.  A K-S test 
reveals a D-statistic of 0.639, or 5.413x10$^{-6}$\%, that stars in the range M0.0-M6.5 and those in the range M7.0-M9.5 are drawn from the same parent distribution.  However, given that we have already 
shown there to be a large difference between early Ms (M0-M3.5), this will bias this probability test.  When we remove all stars earlier than M4 we find a D statistic of 0.544, which relates 
to a probability of only 1.437x10$^{-3}$\% that these are drawn from the same parent population.  This $>$5$\sigma$ result may 
indicate that at temperatures when dust opacity becomes important there is a change in the rotational braking mechanisms and hence the magnetic properties of ultracool dwarfs.  
This might give rise to the flattening trend indicated between M6.5-M9 stars, however 
a more comprehensive study is needed, particularly to decouple the age of these stars by studying the space motion to determine if they are young or old disk stars.  Also the biases of the 
literature surveys are important.  For instance, studies like those of \citet{west09} focus only on selecting inactive, and hence slowly rotating, late type M stars.  In addition, current models 
show that the late M star regime can also be populated by young brown dwarfs.  Finally, this relation also 
suffers from the lack of low $v$~sin~$i$ detections already mentioned above, even more so given the M6.5 detection boundary.  Therefore we expect this result might also become more pronounced 
with further low $v$~sin~$i$ detections at spectral types below M6.5.

\begin{figure}
\vspace{4.0cm}
\hspace{-4.0cm}
\includegraphics{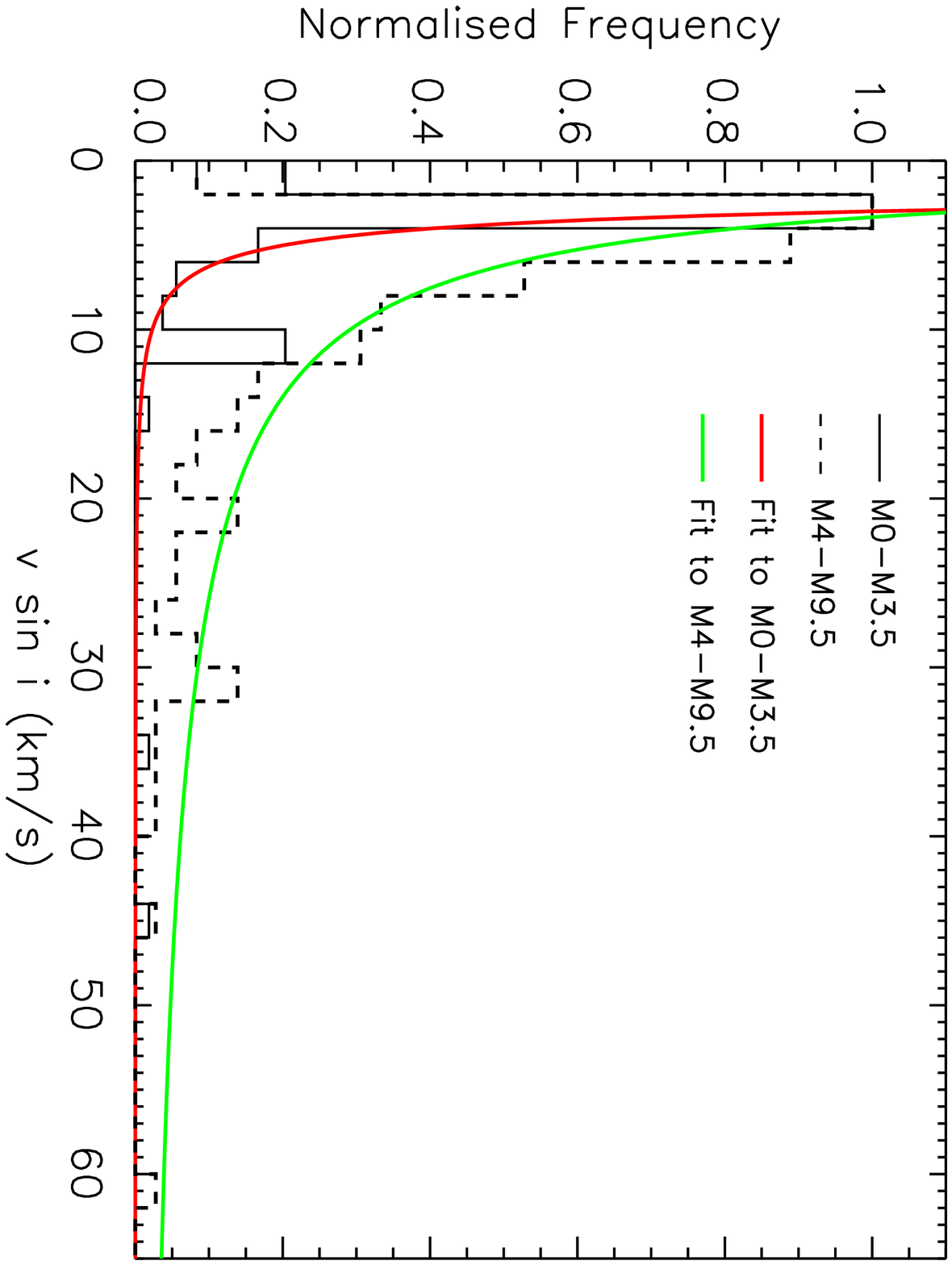}
\vspace{1.2cm}
\caption{Histograms of rotation velocities in this sample and in the literature split by spectral type.  The solid curve represents the stars in the spectral range from M0-M3.5, 
whereas the dashed curve represents the stars between M4-M9.5.  Both samples peak at low rotation velocities of around 
$\sim$3~km~s$^{-1}$, however the bins that contain the later type objects have many more stars with measureable $v$~sin~$i$.  The solid curves are the best fit power laws 
to the data, with the red (dark grey) curve representing the M0-M3.5 bins and the green (light grey) curve representing the M4-M9.5 data.  The changing power law between the two 
spectral regions highlight a possible change in the rotational distribution for fully convective stars.}
\label{vsini_hist}
\end{figure}

The normalized distribution of $v$~sin~$i$s are represented by the histograms in Fig.~\ref{vsini_hist}, where the solid histogram is for all stars in the spectral 
range between M0-M3.5 and the dashed histogram is for all stars in the range M4-M9.5.  These include all values determined in this work combined with those in the   
literature.  It is apparent that both 
distributions peak at low rotation rates ($\sim$3~km~s$^{-1}$), with peak values of 55 and 42 stars respectively.  We find that the total number of $v$~sin~$i$s 
$\le$~10~km~s$^{-1}$ is 198 and these should represent useful stars for future near infrared radial-velocity planet search projects such as PRVS.  
\citet{bouchy01} show that the information content drops by a factor of $\sim$3.5 between rotation velocities of $\sim$2-10~km~s$^{-1}$, making $\ge$~10~km~s$^{-1}$ a reasonable M 
star radial-velocity selection cut.  This sample is still large (124) 
when we include all mid-to-late M stars in the range M3-M9.5 (stars where obtaining optical precision radial-velocities becomes extremely difficult).  
Note the binary systems have been left out of Figs.~\ref{vsini_hist} and \ref{vsini_plot} since the combined luminosities will generate 
inaccurate photometry and therefore inaccurate spectral types.  Also, binary systems like these make radial-velocity exoplanet searches much harder since any small planetary 
signature is masked by the large short period binary velocity, meaning these are not ideal planet search targets for precision radial-velocity programs moving into an unexplored 
parameter space.

Comparing the distributions of both histograms helps to probe the possible changing rotation properties of M stars at the fully convective boundary.  We have employed two 
power law fits to each distribution separately in order to test the changing velocity distribution between these two regimes.  The red (dark grey)\footnote{The colours 
in brackets relate to the printed document, whereas the colours in the text are for the online edition of the article.} curve is fit to the sample of early M dwarfs between M0-M3.5, 
whereas the green (light grey) curve is fit to the mid-to-late M dwarf sample.  The fits are made to the bins by including Poisson uncertainties which are not shown in the 
plot for clarity.  It can be seen that the fit to the early M stars drops much more rapidly than the fit to the later Ms.  The power laws are described by $\delta$N/$\delta$$v$~sin~$i$~$\propto$~$x$$^{-3.13}$ for the early Ms, whereas for the later Ms it is only found to be $\propto$~$x$$^{-1.12}$, highlighting the differing 
steepness of each slope.  The early Ms have a much longer tail than the later Ms due to this faster decay of the distribution and this is also probably an underestimate since we 
included all upper limits in the data to increase the sample size, which included a number of stars from the \citet{stauffer86} sample with detection limits of 10~km~s$^{-1}$.  We note 
that the curvature of these slopes are also affected by the activity lifetimes shown in Table~\ref{tab:sat} since they change with spectral type.

The difference of 2.01 in the exponent between the two spectral samples allows us an insight into the efficiency of the braking mechanism between partially and fully 
convective stars, assuming the larger measured rotation in the later M star sample is not due to increased line blending from increased molecular bands and instrumental 
resolution arguments.  \citet{delfosse98} have shown that the majority of rapidly rotating mid-M stars are members of the young disk population, whereas the older population 
tend to rotate more slowly.  If this is indeed the case, then the braking mechanism is at play in later M stars but the efficiency has dropped across the fully convective boundary.  
As mentioned earlier, the change in the field topology between the partially and fully convective boundary is probably the driving factor which governs the efficiency of the 
wind braking mechanism since \citet{reiners07a} have shown that fully convective stars produce field strengths as strong as partially convective stars.  To better probe 
the braking mechanism in this fashion, in addition to gaining more data, it is necessary to also fold in an age proxy for the sample, and decouple both the young and old 
disk populations to compare these objects.  Along with this a better understanding of the activity lifetimes and how these change with spectral type should be considered.  More 
magnetic field topology studies are required for later M stars which can help to confirm if the later Ms also have axisymetric poloidal fields and add weight to the topology 
argument.  Finally, more detailed testing of changes around the dusty regime ($\sim$M6.5) where stars can also be young brown dwarfs might usefully be investigated.

\begin{sidewaystable*}
\tiny
\caption{Characteristics of M stars in this study.}
\label{tab:vsini}
\begin{tabular}{ccccccccccccc}
\hline
\multicolumn{1}{c}{Star}& \multicolumn{1}{c}{$V$}& \multicolumn{1}{c}{$J$} & \multicolumn{1}{c}{$H$} & 
\multicolumn{1}{c}{$K_{\rm{s}}$} & \multicolumn{1}{c}{Spec Type}& 
\multicolumn{1}{c}{T$_{\rm{EFF}}$~(K)} & \multicolumn{1}{c}{$\pi$~(mas)} & 
\multicolumn{1}{c}{M/M$_{\odot}$} & \multicolumn{1}{c}{[Fe/H]$_{phot}$} & \multicolumn{1}{c}{$v$~sin~$i$~(km~s$^{-1}$)} & 
\multicolumn{1}{c}{FWHM$_{\rm{Tell}}$~(km~s$^{-1}$)} & \multicolumn{1}{c}{H$\alpha$ Emission}  \\ \hline
                                          &       &   &           &              &  &&&&&&&  \\
G121-028            & 14.57&  9.94&  9.31&  9.02& M4.5&           2948        &   54.00$\pm$8.00&      0.200$\pm$0.005&     -- &    3.8$\pm$0.7         &     7.2   &  No*  \\
G180-011$\dagger$            & 13.68&  8.93&  8.26&  8.04& M4.5&           2913        &   92.00$\pm$14.00&      0.184$\pm$0.003&    -- &   21.9$\pm$0.4    &        6.8  &  Yes  \\
GJ1029              & 14.81&  9.49&  8.88&  8.55&   M5.0&           2765        &   79.30$\pm$3.00&     0.158$\pm$0.016&      -- &    4.1$\pm$0.8    &     7.2  &  No  \\
GJ1034              & 15.05& 10.70& 10.17&  9.91&   M4.0&           3035        &   48.10$\pm$4.50&     0.166$\pm$0.025&      -- &    $\le$4.5         &         10.9           &  No*  \\
GJ1055              & 14.86&  9.93&  9.33&  9.07&   M5.0&           2863        &   83.90$\pm$4.00&     0.133$\pm$0.010&      -- &    4.7$\pm$0.7    &              7.5  &  No  \\
GJ1078              & 15.52& 10.70& 10.19&  9.85&   M4.5&           2893        &   48.00$\pm$4.50&     0.158$\pm$0.006&      -- &    7.2$\pm$0.9    &              7.6             &  Yes  \\
GJ1119              & 13.32&  8.60&  8.03&  7.74&   M4.5&           2922        &   96.90$\pm$2.70&     0.200$\pm$0.006&      -- &    4.0$\pm$0.7    &              6.8 &  Yes  \\
GJ1134              & 12.96&  8.49&  8.01&  7.71&   M4.5&           2996        &   96.70$\pm$2.30&     0.212$\pm$0.007&      -- &    4.1$\pm$0.7         &         7.1   &  No*  \\
GJ1182$\dagger$     & 14.30&  9.43&  8.94&  8.62&   M5.0&           2880        &   71.70$\pm$3.40&     0.180$\pm$0.003&      -- &    6.8$\pm$0.5    &              6.5 &  No  \\
GJ1186              & 15.29& 10.58&  9.97&  9.65&   M4.5&           2924        &   53.50$\pm$4.10&     0.157$\pm$0.008&      -- &    3.9$\pm$0.6    &        7.8 &  No  \\
GJ1187              & 15.52& 10.21&  9.64&  9.27&   M5.5&           2767        &   89.00$\pm$4.60&     0.114$\pm$0.006&      -- &   13.6$\pm$0.6    &             6.3 &  Yes  \\
GJ1223              & 14.89&  9.72&  9.19&  8.89&   M5.0&           2802        &   83.50$\pm$3.90&     0.139$\pm$0.002&      -- &   $\le$4.5         &        10.0 &  No*  \\
GJ1250              & 14.88&  9.96&  9.40&  9.08&   M4.5&           2727        &   46.80$\pm$6.20&     0.213$\pm$0.023&   0.03  &   15.7$\pm$0.4    &             8.1  &  Yes    \\
GJ1253              & 14.04&  9.03&  8.48&  8.10&   M5.0&           2842        &  107.50$\pm$3.60&     0.152$\pm$0.002&      -- &   $\le$4.5          &        12.1 &  No/Yes  \\
GJ1268              & 14.94& 10.16&  9.60&  9.30&   M4.5&           2904        &   62.70$\pm$3.60&     0.157$\pm$0.008&      -- &    $\le$4.5    &         11.5                      &  No  \\
GJ2045              & 15.28& 10.21&  9.69&  9.37&   M5.0&           2827        &          --&                --&             -- &    4.6$\pm$0.3    &              7.0                    &  No  \\
GJ3028              & 16.09& 10.28&  9.69&  9.33&   M5.5&           2654        &   79.30$\pm$3.70&     0.115$\pm$0.005&      -- &    $\le$4.5    &       12.9  &  No  \\
GJ3104              & 12.83&  9.14&  8.52&  8.28&   M3.0&           3289        &   41.00$\pm$3.00&     0.386$\pm$0.001&  -0.20 &     4.0$\pm$0.3    &              7.6        &  No*  \\
GJ3128              & 15.61&  9.84&  9.25&  8.93&   M6.0&           2662        &  112.00$\pm$3.20&     0.105$\pm$0.002&      -- &    $\le$4.5    &         12.3            &  No  \\
GJ3147              &    --&  9.98&  9.35&  9.01&   M5.0&     2380$^{*}$        &      --&                    --&            --  &   28.2$\pm$0.7    &              11.1         &  Yes   \\
GJ3153              & 14.78& 10.06&  9.56&  9.20&   M4.5&           2922        &   41.00$\pm$17.00&     0.234$\pm$0.021& 0.00   &   23.3$\pm$0.7    &             7.0  &  Yes   \\
GJ3172              & 15.16& 10.55&  9.97&  9.69&   M4.0&           2954        &   40.20$\pm$4.30&     0.200$\pm$0.001&     --  &    4.9$\pm$0.9    &          7.4                        &  Yes   \\
GJ3181              & 16.86& 10.97& 10.52& 10.19&   M6.0&           2636        &   68.50$\pm$3.50&     0.100$\pm$0.006&      -- &    6.2$\pm$0.8    &           7.1   &   No  \\
GJ3225              & 14.96& 10.12&  9.52&  9.25&   M4.5&           2888        &   60.00$\pm$9.00&     0.164$\pm$0.002&      -- &   21.9$\pm$0.6    &     8.8        &   Yes  \\
GJ3234              & 16.67& 12.05& 11.54& 11.26&   M5.0&           2951        &   34.60$\pm$4.70&     0.129$\pm$0.023&      -- &    $\le$4.5    &         11.9                       &   No  \\
GJ3311              & 16.45& 11.56& 11.06& 10.76&   M6.0&           2874        &   52.10$\pm$0.90&     0.110$\pm$0.019&      -- &    $\le$4.5    &  11.1     &   No  \\
GJ3396              & 14.83&  9.68&  9.12&  8.81&   M5.0&           2807        &   90.00$\pm$15.00&     0.134$\pm$0.003&     -- &   39.6$\pm$1.7    &              7.7                        &   Yes   \\
GJ3421              & 13.30&  8.54&  8.09&  7.78&   M5.0&           2910        &  108.50$\pm$2.10&     0.180$\pm$0.004&      -- &    4.7$\pm$0.6    &                7.4               &   No*   \\
GJ3444$\dagger$              & 15.16& 11.92& 11.38& 11.09&   M6.0&           3519 ?      &   55.00$\pm$4.00&     0.125$\pm$0.057&      -- &    $\le$4.5         &         10.3            &   No   \\
GJ3515              & 15.40& 10.68& 10.08&  9.80&   M4.5&           2922        &   46.00$\pm$17.00&     0.168$\pm$0.005&     -- &    13.1$\pm$0.7    &          7.2   &   Yes   \\
GJ4030              & 13.54&  9.32&  8.63&  8.45&   M3.5&           3079        &   44.80$\pm$4.10&     0.312$\pm$0.020&  -0.04  &    $\le$4.5         &         10.5           &   No*   \\
GJ4108              & 15.37& 10.75& 10.24&  9.97&   M4.5&           2951        &   47.30$\pm$4.10&     0.159$\pm$0.016&      -- &    $\le$4.5         &  10.6      &   No   \\
GJ4345$\dagger$    & 16.10& 11.09& 10.48& 10.17&   M5.0&           2842        &   41.60$\pm$3.20&     0.152$\pm$0.001&      -- &   9.8$\pm$1.1    &              6.9             &   Yes   \\
GJ4385              & 16.10& 10.87& 10.26&  9.93&   M5.0&           2787        &   51.90$\pm$0.90&     0.136$\pm$0.002&      -- &    $\le$4.5    &              12.0                        &   No   \\
Gl359               & 14.20&  9.63&  9.09&  8.81&   M4.5&           2966        &   79.00$\pm$3.80&     0.162$\pm$0.016&       -- &    3.2$\le$1.1         &         7.0                        &   No   \\
Gl585               & 13.68&  9.11&  8.62&  8.30&   M4.5&           2966        &   85.10$\pm$2.90&     0.184$\pm$0.009&       -- &    3.1$\pm$0.5         &  6.4       &   No*   \\
LHS252              & 15.05&  9.61&  9.00&  8.67&   M5.5&           2737        &   99.80$\pm$3.40&     0.126$\pm$0.003&      -- &    4.0$\pm$0.5    &          7.4   &   Yes   \\
LHS265              & 15.11& 10.26&  9.70&  9.40&   M4.5&           2885        &   64.40$\pm$4.00&     0.147$\pm$0.008&      -- &    3.9$\pm$0.8         &      7.4   &   No   \\
LHS302              & 15.14& 10.29&  9.74&  9.46&   M5.0&           2885        &   57.20$\pm$3.70&     0.158$\pm$0.006&      -- &    $\le$2.5    &      7.0         &   No   \\
LHS1785             & 14.54& 10.04&  9.51&  9.18&   M4.5&           2987        &   58.80$\pm$1.90&     0.180$\pm$0.012&     -- &    4.5$\pm$0.6    &    6.8            &   No   \\
LHS1809             & 14.48&  9.35&  8.77&  8.44&   M5.0&           2812        &  107.70$\pm$2.60&     0.133$\pm$0.003&     -- &    4.3$\pm$1.2         &  6.4         &   No   \\
LHS1857             & 14.22&  9.79&  9.31&  8.99&   M4.5&           3009        &   54.70$\pm$2.40&     0.209$\pm$0.009&     -- &    4.0$\pm$0.5         &         7.3            &   No   \\
LHS1950             & 14.75&  9.97&  9.40&  9.09&   M4.5&           2904        &   62.70$\pm$3.10&     0.169$\pm$0.002&     -- &   $\le$2.5          & 7.3           &   No   \\
LHS2090             &    --&  9.44&  8.84&  8.44&   M6.5&    2753$^{**}$        &            --&        --&                 --  &   20.0$\pm$0.6     &              6.7                       &   Yes    \\
LHS2206             & 14.05&  9.21&  8.60&  8.33&   M4.5&           2888        &  108.39$\pm$2.30&     0.143$\pm$0.009&    --  &   16.5$\pm$0.4    &            7.1  &   Yes    \\
LHS2320             & 14.40&  9.83&  9.28&  8.94&   M5.0&           2966        &   46.00$\pm$8.00&     0.239$\pm$0.016& -0.03  &   19.1$\pm$0.2    &           6.3  &   Yes    \\
LHS2930$\dagger$   &    --& 10.79& 10.14&  9.79&   M6.5&    2906$^{**}$ ?       &            --&              --&       --      &  18.7$\pm$1.5     &              8.2            &    Yes     \\
LHS3075             & 14.17&  9.59&  9.02&  8.72&   M4.5&           2963        &   51.10$\pm$4.40&     0.239$\pm$0.015& -0.04   &  $\le$2.5         & 5.8               &    No   \\
LP205-49            & 18.71& 14.70& 14.19& 13.90&   M3.5&         3156          &      --         &            --      &   --    &  45.4$\pm$2.0    &                      7.7   &    Yes      \\

\hline
\end{tabular}

\medskip
The $\dagger$ symbol relates to deconvolved profiles that show either weak evidence for a possible binary component, but not significant enough to be included in the binary table, however it may affect the 
deconvolved profile enough to give rise to a larger rotation velocity than the star has, or stars where the deconvoluted profile has low S/N and the construction is of 
lower quality, particularly in the wings, which again can give rise to an inaccurate value for the $v$~sin~$i$.  The question marks after the effective temperature are used to flag highly 
suspect values.  The asterisks in the H$\alpha$ column represent stars with significant absorption.  The uncertainties shown for the $v$~sin~$i$s represent the formal uncertainties 
on each parameter fit.  However, as explained in the text, the actual uncertainties for high S/N spectra of objects rotating $\le$20~km~s$^{-1}$ is $\pm$1~km~s$^{-1}$, increasing to 
around $\pm$10-20\% for the fastest rotating objects.  

\end{sidewaystable*}

\begin{table*}
\footnotesize
\caption{Table of $v$~sin~$i$s taken from the literature and used in this work.  Columns are the same as Table~\ref{tab:vsini} except without the telluric widths in column 15.}
\label{tab:lit_vsini}
\begin{tabular}{ccccccccccccc}
\hline
\multicolumn{1}{c}{Star}& \multicolumn{1}{c}{$V$}& \multicolumn{1}{c}{$J$} & \multicolumn{1}{c}{$H$} & 
\multicolumn{1}{c}{$K_{\rm{s}}$} & \multicolumn{1}{c}{Spec}& 
\multicolumn{1}{c}{T$_{\rm{EFF}}$} & \multicolumn{1}{c}{$\pi$} & \multicolumn{1}{c}{M/M$_{\odot}$} & \multicolumn{1}{c}{[Fe/H]} & \multicolumn{1}{c}{$v$~sin~$i$} & 
\multicolumn{1}{c}{Source}\\

\multicolumn{1}{c}{}& \multicolumn{1}{c}{}& \multicolumn{1}{c}{} & \multicolumn{1}{c}{} & 
\multicolumn{1}{c}{} & \multicolumn{1}{c}{Type}& 
\multicolumn{1}{c}{(K)} & \multicolumn{1}{c}{(mas)} & 
\multicolumn{1}{c}{} & \multicolumn{1}{c}{} & \multicolumn{1}{c}{(km~s$^{-1}$)} & \multicolumn{1}{c}{} \\ \hline

2MASS-J1242081+290027         &     --  &     16.42  &     16.03  &     15.45  &           M8.0  &        --  &       --  &       --  &       --  &     7.0 & (f) \\
2MASS-J1254012+250002         &     --  &     15.29  &     14.80  &     14.81  &           M7.5  &        --  &       --  &       --  &       --  &    13.0 & (f) \\
2MASS-J1255583+275947         &     --  &     15.42  &     14.73  &     14.58  &           M7.5  &        --  &       --  &       --  &       --  &     9.0 & (f) \\
AD Leo                        & 9.43    &   5.45  & 4.84    &   4.59  &          M4.5  &     3157  & 213.00$\pm$4.00  & 0.390$\pm$0.032  &  0.04  &    3.0 & (j) \\
BRI0021-0214         &  19.60  &  11.99  &  11.08  &  10.48  &           M9.5  &     2092  &    80.00$\pm$3.40  &    0.103$\pm$0.031  &       --  &    34.0 & (f) \\
BRI1222-1222         &     --  &  12.57  &  11.82  &  11.35  &           M9.0  &       --  &       --  &      --  &       --  &     8.0 & (f) \\
CTI0126+57.5         &     --  &     --  &     --  &     --  &           M9.0  &       --  &       --  &      --  &       --  &    11.1 & (f) \\
CTI0156+28         &     --  &     14.47  &     13.84  &     13.54  &           M6.5  &       --  &        --  &       --  &       --  &     9.0  & (f) \\
CTI115638.4+28         &     --  &     14.32  &     13.72  &     13.34  &           M7.0  &       --  &        --  &       --  &       --  &    10.5 & (f) \\
CTI1539+28         &     --  &     15.49  &     15.00  &     14.58  &           M6.5  &       --  &            --  &       --  &       --  &     7.0 & (f) \\
CTI1747+28         &     --  &     15.50  &     15.04  &     14.52  &           M6.5  &       --  &            --  &       --  &       --  &    45.0 & (f) \\
CTI2332+27         &     --  &     15.71  &     15.34  &     14.86  &           M6.0  &       --  &            --  &       --  &       --  &    25.0 & (f) \\
DENIS-J0021-4243          &     --  &     15.80  &     15.39  &     15.24  &           M9.5  &       --  &       --  &       --  &       --  &    17.5 & (f) \\
DENIS-1048-3955        &     --  &      14.60  &     14.23  &     14.19  &           M8.0  &       --  &       --  &       --  &       --  &    25.0 & (h) \\
DENIS-J1207+0059         &     --  &      10.38  &     10.13  &     10.06  &           M9.0  &       --  &       --  &       --  &       --  &    10.0 & (f) \\
DENIS-J1431596-195321         &     --  &     15.34  &     14.73  &     14.45  &           M9.0  &           --  &       --  &       --  &       --  &    37.1 & (g) \\
ESO207-61         &  20.99  &  13.23  &  12.54  &  12.06  &           M8.0  &       --  &    70.00$\pm$4.00  &    0.139$\pm$0.088  &       --  &    10.0 & (f) \\
G087-09B         &     --  &     10.40  &     9.79  &     9.76  &           M4.0  &       --  &          --  &       --  &       --  &     6.0 & (i) \\
G089-032         &     --  &     8.18  &     7.61  &     7.28  &           M5.0  &       --  &          --  &       --  &       --  &     7.9 & (f) \\
G099-049         &     --  &     6.91  &     6.31  &     6.04  &           M4.0  &       --  &          --  &       --  &       --  &     7.4 & (f) \\
G165-08          &   12.19 &    7.56   &  7.00    &     6.72   &          M4.0   &    2951   & 126.00       &   0.242$\pm$0.018    &      -0.02     &  55.5 & (f) \\
G188-38         &  11.98  &     7.64  &     7.04  &     6.78  &           M4.0  &       --  &          --  &       --  &       --  &    29.4 & (f) \\
GJ65A         &  12.57  &     --  &     --  &     --  &           M5.5  &       --  &            --  &       --  &       --  &    31.5 & (i) \\
GJ65B         &  12.52  &     --  &     --  &     --  &           M6.0  &       --  &           --  &       --  &       --  &    29.5 & (i) \\
GJ166C         &     --  &   6.75  &   6.28  &   5.96  &           M4.5  &       --  &           --  &       --  &       --  &     5.0 & (i) \\
GJ630.1A         &     --  &   8.50  &   8.04  &   7.80  &           M4.5  &     2887$^{*}$  &        --  &       --  &       --  &    27.5 & (i) \\
GJ699         &     --  &    5.24  &    4.83  &    4.52  &           M4.0  &       --  &       --  &       --  &       --  &     $\le$2.8 & (f) \\
GJ725A         &   8.91  &   5.19  &   4.74  &   4.41  &           M3.0  &     3276  &   286.10$\pm$1.80  &    0.342$\pm$0.020  &    -0.35  &     $\le$5.0 & (i) \\
GJ725B         &   9.69  &   5.72  &   5.20  &   4.98  &           M3.5  &     3172  &   286.10$\pm$1.80  &    0.266$\pm$0.025  &    -0.38  &     $\le$7.0 & (i) \\
GJ896A         &  10.35  &   6.16  &   5.57  &   5.31  &           M3.5  &     3090  &   151.90$\pm$3.70  &    0.378$\pm$0.042  &     0.11  &    10.0 & (i) \\
GJ896B         &     --  &   7.10  &   6.56  &   6.26  &           M4.5  &       --  &             --  &       --  &       --  &    15.0 & (i) \\
GJ1002         &  13.73  &    8.32  &     7.79  &     7.44  &           M5.5  &       --  &           --  &       --  &       --  &     $\le$3.0 & (f) \\
GJ1057         &  13.78  &   8.77  &   8.21  &   7.80  &           M5.0  &     2706  &   120.00$\pm$3.50  &    0.155$\pm$0.005  &       --  &     $\le$2.2 & (f) \\
GJ1093         &  14.83  &   9.16  &   8.55  &   8.23  &           M5.0  &       --  &          --  &       --  &       --  &     $\le$2.8 & (f) \\
GJ1105         &  12.04  &   7.73  &   7.13  &   6.88  &           M3.5  &       --  &          --  &       --  &       --  &     $\le$2.0 & (c) \\
GJ1111         &  14.81  &   8.24  &   7.62  &   7.26  &           M6.5  &       --  &          --  &       --  &       --  &    11.0 & (f) \\
GJ1111         &  14.81  &   8.24  &   7.62  &   7.26  &           M6.5  &       --  &          --  &       --  &       --  &     8.1 & (c) \\
GJ1151         &  13.26  &   8.49  &   7.95  &   7.61  &           M4.5  &     2767  &   120.00$\pm$2.90  &    0.174$\pm$0.001  &       --  &     $\le$4.1 & (f) \\
GJ1154A         &  14.11  &  8.46  &   7.86  &   7.54  &           M5.0  &       --  &           --  &       --  &       --  &     5.2 & (f) \\
GJ1156         &  13.79  &   8.52  &   7.88  &   7.54  &           M5.0  &     2632  &   150.00$\pm$3.00  &    0.139$\pm$0.006  &       --  &     9.2 & (f) \\
GJ1224         &  13.64  &   8.64  &   8.09  &   7.81  &           M4.5  &     2711  &   130.00$\pm$3.70  &    0.148$\pm$0.004  &       --  &     $\le$5.6 & (f) \\
GJ1227         &  13.40  &   8.64  &   8.05  &   7.71  &           M4.5  &     2910  &   120.00$\pm$2.20  &    0.167$\pm$0.002  &       --  &     $\le$2.3 & (f) \\
GJ1230B         &  14.40  &   8.86  &   8.03  &   7.73  &           M5.0  &     2714  &   120.00$\pm$7.20  &    0.147$\pm$0.024  &       --  &    $\le$ 7.1 & (f) \\
GJ1245A         &  13.41  &   7.79  &   7.19  &   6.82  &           M5.5  &     2695  &   220.00$\pm$1.00  &    0.129$\pm$0.010  &       --  &    22.5 & (f) \\
GJ1245B         &  13.99  &   8.27  &   7.73  &   7.36  &           M5.5  &     2673  &   220.00$\pm$1.00  &    0.108$\pm$0.001  &       --  &     6.8 & (f) \\
GJ1286         &  14.68  &   9.15  &   8.51  &   8.15  &           M5.5  &     2716  &   140.00$\pm$3.50  &    0.116$\pm$0.001  &       --  &     $\le$5.7 & (f) \\
GJ1289         &  12.67  &   8.11  &   7.45  &   7.20  &           M4.0  &     2969  &   120.00$\pm$2.90  &    0.209$\pm$0.004  &       --  &     $\le$2.6 & (f) \\
GJ2005         &     --  &   9.25  &   8.55  &   8.24  &           M6.0  &       --  &           --  &       --  &       --  &     9.0 & (f) \\
GJ2066         &  10.05  &   6.63  &   6.04  &   5.74  &           M2.0  &     3420  &   114.00$\pm$3.90  &    0.447$\pm$0.005  &    -0.22  &     $\le$2.7 & (c) \\
GJ2069B         &  13.40  &   8.62  &   8.05  &   7.72  &           M4.0  &     2904  &            --  &       --  &       --  &     6.5 & (f) \\
GJ2097         &  12.54  &    --  &     --  &     --  &           M1.5  &       --  &            --  &       --  &       --  &     $\le$3.7 & (c) \\
GJ3136         &  12.47  &   8.42  &   7.81  &   7.55  &           M5.0  &     3141  &    70.00$\pm$4.00  &    0.312$\pm$0.006  &    -0.14  &    30.0 & (e) \\
GJ3304         &  12.51  &   8.17  &   7.62  &   7.30  &           M4.0  &     3038  &   100.00$\pm$7.00  &    0.243$\pm$0.002  &    -0.14  &    30.0 &  (e) \\
GJ3323         &  12.16  &   7.62  &   7.07  &   6.71  &           M4.0  &     2975  &   163.00$\pm$6.00  &    0.196$\pm$0.002  &       --  & $\le$3.2 & (f)  \\
GJ3378         &  11.71  &   7.47  &   6.95  &   6.62  &           M3.5  &     3072  &   132.20$\pm$2.90  &    0.255$\pm$0.002  &    -0.18  & $\le$2.7 & (c) \\
GJ3482B         &  11.20  &   7.34  &   6.76  &   6.50  &           M2.5  &     3216  &    59.00$\pm$7.00  &    0.532$\pm$0.066  &       --  &    35.0 & (e) \\
GJ3622         &  15.60  &   8.86  &   8.26  &   7.90  &           M6.0  &       --  &   220.00$\pm$3.60  &         --  &  -- &   3.0 & (f) \\
GJ3828B         &     --  &  13.09  &  12.53  &  12.09  &           M6.0  &       --  &            --  &       --  &       --  &     9.0 & (f) \\
GJ3877         &  17.05  &   9.97  &   9.31  &   8.93  &           M7.0  &     2393  &             --  &       --  &       --  &     8.0 & (f) \\

\hline
\end{tabular}
\medskip
\end{table*}

\begin{table*}
\footnotesize
\begin{tabular}{ccccccccccccc}
\hline

GJ4281         &  17.14  &  10.77  &  10.22  &   9.81  &           M6.5  &     2536  &    90.00$\pm$4.90  &    0.093$\pm$0.005  &       --  &     7.0 & (f) \\
Gl14         &   8.94  &   6.39  &   5.75  &   5.56  &           M0.5  &     4021  &    70.00$\pm$3.90  &    0.721$\pm$0.029  &    -0.30  &     $\le$2.6 & (b) \\
Gl15A         &   8.07  &    5.25  &   4.48  &   4.02  &           M2.0  &       --  &            --  &       --  &       --  &     2.9 & (c)  \\
Gl15B         &  11.04  &    6.79  &   6.19  &   5.95  &           M6.0  &       --  &             --  &       --  &       --  &     $\le$3.1 & (c) \\
Gl26         &  11.06  &    7.45  &   6.86  &   6.61  &           M4.0  &       --  &            --  &       --  &       --  &     $\le$2.9 & (b) \\
Gl48         &   9.96  &   6.30  &   5.70  &   5.42  &           M3.0  &     3303  &   115.50$\pm$3.70  &    0.480$\pm$0.030  &     0.04  &     $\le$2.4 & (c) \\
Gl49         &   9.56  &   6.23  &   5.58  &   5.37  &           M1.5  &     3468  &             --  &       --  &       --  &     $\le$3.4 & (c) \\
Gl70         &  10.96  &   7.37  &   6.81  &   6.49  &           M2.0  &     3335  &    90.00$\pm$5.10  &    0.403$\pm$0.003  &    -0.22  &     $\le$3.0 & (c) \\
Gl82         &  12.04  &   7.79  &   7.22  &   6.94  &           M4.0  &     3069  &    78.50$\pm$4.90  &    0.349$\pm$0.035  &     0.06  &    15.0 & (a) \\
Gl83.1         &  12.26  &  7.51  &   6.97  &   6.65  &           M4.5  &       --  &            --  &       --  &       --  &     3.8 & (f) \\
Gl87         &  10.03  &   6.83  &   6.32  &   6.06  &           M2.5  &     3542  &    86.70$\pm$7.40  &    0.512$\pm$0.022  &    -0.35  &     $\le$1.1 &(b) \\
Gl105B         &  11.66  &   7.33  &   6.79  &   6.55  &           M4.5  &     3042  &   129.40$\pm$4.30  &    0.266$\pm$0.003  &    -0.14  &     $\le$2.4 & (f) \\
Gl109         &  10.58  &   6.75  &   6.20  &   5.94  &           M3.5  &     3228  &   130.00$\pm$4.20  &    0.359$\pm$0.002  &    -0.21  &     $\le$2.8 & (c) \\
Gl157B         &  11.48  &   7.77  &   7.16  &   6.90  &           M3.0  &     3280  &    70.00$\pm$5.20  &    0.416$\pm$0.012  &    -0.10  &    $\le$10.0 & (a) \\
Gl169.1A         &  11.08  &   6.62  &   6.01  &   5.69  &           M4.0  &     3000  &   180.00$\pm$0.80  &    0.271$\pm$0.025  &     0.01  &     1.9 & (f) \\
Gl205         &   7.92  &    5.00  &    4.15  &    4.04  &           M1.5  &       --  &            --  &       --  &       --  &     1.0 & (j) \\
Gl206         &  11.48  &   7.42  &   6.88  &   6.53  &           M4.0  &     3137  &    70.00$\pm$3.90  &    0.452$\pm$0.063  &       0.27  &    10.0 & (a)\\
Gl207.1         &  11.52  &   7.76  &   7.15  &   6.83  &           M2.5  &     3258  &    60.00$\pm$5.20  &    0.473$\pm$0.044  &     0.15  &    10.0 & (a) \\
Gl213         &  11.48  &    7.12  &    6.63  &   6.39  &           M4.0  &       --  &              --  &       --  &       --  &     $\le$2.9 & (f) \\
Gl229A        &  8.14   &    5.10  & 4.39   &    4.17   &           M1/M2 &   3571   &  173.19$\pm$1.12 &     0.583$\pm$0.005    & -0.09  &     1.0 & (j) \\
Gl232         &  13.06  &   8.66  &   8.16  &   7.89  &           M4.5  &     3019  &   120.00$\pm$2.30  &    0.168$\pm$0.023  &       --  &     $\le$3.1 & (f) \\
Gl234A         &  11.10  &   6.38  &   5.75  &   5.46  &           M4.5  &     2923  &   240.00$\pm$3.70  &    0.224$\pm$0.020  &     0.00  &     6.0 & (f) \\
Gl251         &   9.89  &   6.10  &   5.53  &   5.26  &           M4.0  &     3245  &   170.00$\pm$3.20  &    0.372$\pm$0.002  &    -0.18  &     $\le$2.4 & (c) \\
Gl268.3         &  10.83  &   7.01  &   6.44  &   6.17  &           M0.0  &     3232  &   126.00$\pm$5.00  &    0.337$\pm$0.007  &    -0.25  &     $\le$2.5 & (d) \\
Gl273         &   9.89  &   5.71  &   5.22  &   4.83  &           M3.5  &     3093  &   264.40$\pm$2.00  &    0.286$\pm$0.007  &    -0.12  &     0.0 & (j) \\
Gl277A         &  11.87  &   6.77  &   6.18  &   5.90  &           M3.5  &     2819  &    90.00$\pm$2.50  &    0.413$\pm$0.142  &      --  &    $\le$10.0 & (a) \\
Gl277B         &  11.79  &   7.57  &   6.99  &   6.74  &           M4.5  &     3079  &    90.00$\pm$2.50  &    0.338$\pm$0.026  &     0.00  &    $\le$10.0 & (a) \\
Gl285         &  11.12  &   6.58  &   6.01  &   5.67  &           M4.5  &     2975  &   170.00$\pm$4.40  &    0.284$\pm$0.035  &     0.07  &     6.5 & (f) \\
Gl299A         &  12.83  &   8.42  &   7.93  &   7.64  &           M4.5  &     3036  &   148.00$\pm$2.60  &    0.155$\pm$0.025  &       --  &     3.0 & (f) \\
Gl338A         &   7.64  &   4.89  &   3.99  &   3.96  &           M0.0  &     3851  &   162.50$\pm$2.00  &    0.660$\pm$0.011  &    -0.17  &     2.9 & (c) \\
Gl338B         &   7.74  &   4.78  &   4.04  &   4.13  &           M0.0  &     3695  &   162.50$\pm$2.00  &    0.634$\pm$0.028  &    -0.35  &     2.8 & (c) \\
Gl362         &  11.36  &   7.33  &   6.73  &   6.44  &           M3.0  &     3148  &    80.00$\pm$3.50  &    0.423$\pm$0.048  &     0.16  &    $\le$10.0 & (a) \\
Gl369         &  10.00  &   6.99  &   6.40  &   6.12  &           M2.0  &     3661  &    84.80$\pm$7.60  &    0.517$\pm$0.031  &    -0.42  &     5.0  & (b) \\
Gl382         &   9.26  &   5.89  &   5.26  &   4.98  &           M2.0  &     3446  &   120.00$\pm$5.90  &    0.556$\pm$0.027  &     0.07  &     1.3 & (j) \\
Gl393         &   9.63  &   6.18  &   5.61  &   5.28  &           M2.5  &     3404  &   130.00$\pm$5.10  &    0.473$\pm$0.008  &    -0.12  &     $\le$1.1 & (j) \\
Gl402         &  11.66  &   7.32  &   6.71  &   6.34  &           M5.0  &     3038  &   145.10$\pm$4.80  &    0.255$\pm$0.014  &    -0.06  &     $\le$2.3 & (f) \\
Gl406         &  13.54  &   7.09  &   6.48  &   6.04  &           M6.0  &     2520  &   418.30$\pm$2.50  &    0.101$\pm$0.005  &       --  &     $\le$3.0 & (f) \\
Gl406         &  13.54  &   7.09  &   6.48  &   6.08  &           M6.0  &     2520  &   418.30$\pm$2.50  &    0.101$\pm$0.005  &       --  &     $\le$2.9 & (c) \\
Gl408         &  10.03  &   6.31  &   5.76  &   5.47  &           M3.0  &     3276  &   140.00$\pm$4.30  &    0.406$\pm$0.007  &    -0.14  &     $\le$2.3 & (c) \\
Gl411         &   7.49  &   4.20  &   3.64  &   3.22  &           M2.0  &     3490  &   397.30$\pm$1.80  &    0.421$\pm$0.017  &    -0.33  &     $\le$2.9 & (c) \\
Gl412A         &   8.68  &   5.54  &   5.00  &   4.77  &           M2.0  &       --  &             --  &       --  &       --  &     $\le$3.0 & (c) \\
Gl412B         &  14.45  &   8.74  &   8.18  &   7.84  &           M6.0  &       --  &             --  &       --  &       --  &     7.7 & (f) \\
Gl414B         &   9.95  &   6.59  &   5.97  &   5.70  &           M2.0  &     3451  &    70.00$\pm$3.60  &    0.636$\pm$0.050  &      --  &     $\le$3.2 & (b) \\
Gl424         &   9.32  &    6.31  &   5.73  &   5.53  &           M1.0  &       --  &            --  &       --  &       --  &     $\le$2.9 & (c) \\
Gl436         &  10.68  &   6.90  &   6.32  &   6.07  &           M3.5  &       --  &             --  &       --  &       --  &     $\le$1.0 & (b) \\
Gl445         &  10.78  &   6.72  &   6.22  &   5.93  &           M4.0  &     3137  &   190.00$\pm$6.00  &    0.254$\pm$0.018  &    -0.31  &     $\le$2.0 & (c) \\
Gl447         &  11.08  &   6.51  &   5.95  &   5.62  &           M4.5  &     2966  &   300.00$\pm$1.70  &    0.179$\pm$0.007  &       --  &     $\le$2.0 & (f) \\
Gl450         &   9.78  &   6.42  &   5.83  &   5.59  &           M1.0  &     3451  &   110.00$\pm$5.70  &    0.491$\pm$0.004  &    -0.21  &     $\le$3.3 & (c) \\
Gl459.3         &  10.62  &   7.67  &   6.97  &   6.77  &           M2.0  &     3702  &    50.00$\pm$1.60  &    0.604$\pm$0.008  &    -0.18  &     $\le$2.8 & (b) \\
Gl461AB         &   9.19  &   6.86  &   6.22  &   6.01  &           M0.0  &       --  &             --  &       --  &       --  &      $\le$2.5 & (b) \\
Gl464         &  10.43  &   7.48  &   6.86  &   6.60  &           M2.0  &     3702  &    49.30$\pm$4.10  &    0.641$\pm$0.001  &    -0.08  &     $\le$2.4 & (b) \\
Gl480         &  11.52  &   7.58  &   6.94  &   6.66  &           M4.0  &     3183  &    80.00$\pm$6.10  &    0.394$\pm$0.031  &     0.04  &     $\le$0.8 & (b) \\
Gl486         &  11.40  &   7.20  &   6.67  &   6.33  &           M4.0  &     3086  &   120.00$\pm$4.00  &    0.311$\pm$0.017  &    -0.06  &     $\le$2.0 & (c) \\
Gl487         &  10.92  &   6.88  &   6.23  &   6.05  &           M3.0  &     3145  &   116.20$\pm$0.50  &    0.365$\pm$0.021  &    -0.04  &    10.0 & (a) \\
Gl490A         &  10.50  &   7.40  &   6.73  &   6.52  &           M0.0  &     3603  &    50.00$\pm$3.60  &    0.639$\pm$0.018  &     0.07  &     8.0 & (a) \\
Gl490B         &  13.16  &   8.87  &   8.28  &   7.99  &           M4.0  &     3055  &    50.00$\pm$3.60  &    0.336$\pm$0.036  &     0.07  &    10.0 & (a) \\
Gl493.1         &  13.37  &   8.55  &   7.97  &   7.63  &           M5.0  &     2893  &   120.00$\pm$3.50  &    0.170$\pm$0.002  &       --  &    16.8 & (f) \\
Gl494         &   9.72  &   6.44  &   5.79  &   5.55  &           M1.5  &     3496  &    90.00$\pm$5.80  &    0.572$\pm$0.020  &     0.03  &    10.0 & (a) \\
Gl507.1         &  10.66  &   7.27  &   6.64  &   6.36  &           M2.0  &     3435  &    50.00$\pm$7.50  &    0.645$\pm$0.058  &      --  &     $\le$2.6 & (b) \\
Gl514         &   9.04  &   5.90  &   5.30  &   5.01  &           M1.0  &     3578  &   138.70$\pm$2.90  &    0.514$\pm$0.014  &    -0.28  &     1.3 & (j) \\
Gl521         &  10.26  &   7.05  &   6.51  &   6.26  &           M2.0  &     3536  &    80.00$\pm$5.30  &    0.506$\pm$0.021  &    -0.34  &     $\le$2.0 & (b) \\
Gl526         &   8.46  &   5.18  &   4.78  &   4.42  &           M4.0  &       --  &             --  &       --  &       --  &     1.4 & (j) \\
Gl552         &  10.68  &   7.23  &   6.61  &   6.36  &           M2.5  &     3404  &    70.00$\pm$4.40  &    0.517$\pm$0.019  &    -0.01  &     $\le$1.2 & (b) \\
Gl555         &  11.35  &   6.84  &   6.26  &   5.91  &           M4.0  &     2984  &   160.00$\pm$7.90  &    0.273$\pm$0.029  &     0.03  &     2.7 & (f) \\
Gl569AB         &  10.20  &   6.63  &   5.99  &   5.74  &           M2.5  &     3345  &    95.60$\pm$1.40  &    0.499$\pm$0.028  &     0.05  &     $\le$3.8 & (b) \\

\hline
\end{tabular}
\medskip
\end{table*}

\begin{table*}
\footnotesize
\begin{tabular}{ccccccccccccc}
\hline

Gl570.2         &  11.08  &   8.44  &   7.82  &   7.64  &           M2.0  &     3942  &    70.00$\pm$5.00  &    0.378$\pm$0.108  &    -1.21  &     $\le$2.5 & (b) \\
Gl570B         &   8.10  &    4.55  &   3.91  &   3.80  &           M2.0  &       --  &             --  &       --  &       --  &     $\le$2.9 & (b) \\
Gl581         &  10.56  &   6.71  &   6.09  &   5.81  &           M5.0  &     3220  &   160.00$\pm$5.60  &    0.313$\pm$0.007  &    -0.24  &     $\le$2.1 & (c) \\
Gl623         &  10.27  &   6.64  &   6.14  &   5.92  &           M3.0  &       --  &              --  &       --  &       --  &     $\le$2.9 & (c) \\
Gl625         &  10.17  &   6.61  &   6.06  &   5.81  &           M2.0  &     3350  &   150.00$\pm$2.50  &    0.349$\pm$0.030  &    -0.44  &     $\le$3.4 & (c) \\
Gl628         &  10.12  &   5.95  &   5.37  &   5.04  &           M3.5  &     3097  &   240.00$\pm$4.20  &    0.286$\pm$0.009  &    -0.11  &     1.1 & (j) \\
Gl643         &  11.70  &   7.55  &   7.06  &   6.70  &           M4.0  &     3104  &   171.90$\pm$7.30  &    0.201$\pm$0.023  &       --  &     $\le$2.7 & (c) \\
Gl649         &   9.62  &   6.45  &   5.86  &   5.60  &           M2.0  &     3560  &   100.00$\pm$4.30  &    0.534$\pm$0.009  &    -0.23  &     $\le$1.9 & (b) \\
Gl654         &  10.07  &   6.78  &   6.19  &   5.95  &           M3.5  &     3490  &   100.00$\pm$7.80  &    0.471$\pm$0.018  &    -0.33  &     $\le$1.1 & (b) \\
Gl669A         &  11.42  &   7.27  &   6.71  &   6.39  &           M4.0  &     3104  &    93.30$\pm$1.90  &    0.376$\pm$0.040  &     0.09  &    $\le$10.0 & (a) \\
Gl669B         &  13.02  &   7.27  &   6.71  &   6.39  &           M5.0  &     2667  &    93.30$\pm$1.90  &    0.311$\pm$0.132  &       --  &    $\le$10.0 & (a) \\
Gl686         &   9.62  &   6.36  &   5.79  &   5.55  &           M1.0  &     3507  &   130.00$\pm$3.80  &    0.447$\pm$0.030  &    -0.44  &     $\le$5.0 & (c) \\
Gl687         &   9.15  &   5.34  &   4.77  &   4.53  &           M3.5  &     3237  &   212.70$\pm$2.00  &    0.407$\pm$0.013  &    -0.09  &     $\le$2.8 & (c) \\
Gl694         &  10.50  &   6.81  &   6.22  &   5.93  &           M3.5  &     3289  &   100.00$\pm$4.10  &    0.447$\pm$0.022  &    -0.02  &     $\le$1.4 & (b) \\
Gl701         &   9.37  &   6.16  &   5.57  &   5.28  &           M2.0  &     3536  &   130.00$\pm$4.30  &    0.489$\pm$0.016  &    -0.30  &     $\le$3.5 & (c) \\
Gl720A         &   9.84  &   6.88  &   6.26  &   6.06  &           M2.0  &     3695  &    67.80$\pm$2.10  &    0.617$\pm$0.012  &    -0.21  &     $\le$1.5 & (b) \\
Gl735         &  10.20  &   6.31  &   5.68  &   5.40  &           M3.0  &     3203  &    90.00$\pm$2.70  &    0.557$\pm$0.087  &       --  &    $\le$10.0 & (a) \\
Gl745A         &  10.84  &   7.30  &   6.73  &   6.50  &           M2.0  &     3359  &   110.00$\pm$4.10  &    0.348$\pm$0.032  &    -0.46  &     $\le$3.0 & (c) \\
Gl745B         &  10.77  &   7.28  &   6.75  &   6.50  &           M2.0  &     3384  &   110.00$\pm$4.10  &    0.352$\pm$0.037  &    -0.50  &     2.8 & (c) \\
Gl748         &  11.06  &   7.09  &   6.57  &   6.27  &           M4.0  &     3172  &   100.00$\pm$2.40  &    0.384$\pm$0.021  &    -0.04  &     4.6 & (b) \\
Gl752A         &   9.13  &   5.58  &   4.93  &   4.64  &           M3.5  &     3354  &   176.70$\pm$2.40  &    0.460$\pm$0.018  &    -0.04  &     $\le$2.6 & (c) \\
Gl781         &  11.97  &   8.83  &   8.35  &   8.09  &           M3.0  &     3578  &    59.90$\pm$2.00  &    0.338$\pm$0.074  &    -0.87  &    15.0 & (a) \\
Gl791.2         &  13.04  &   8.23  &   7.67  &   7.28  &           M6.0  &     2896  &   110.00$\pm$1.90  &    0.210$\pm$0.020  &     0.01  &    32.0 & (f) \\
Gl793         &  10.44  &   6.74  &   6.14  &   5.91  &           M3.0  &     3284  &   120.00$\pm$3.10  &    0.393$\pm$0.001  &    -0.20  &     $\le$3.2 & (c) \\
Gl806         &  10.84  &   7.33  &   6.77  &   6.51  &           M3.0  &     3374  &    80.00$\pm$2.20  &    0.446$\pm$0.004  &    -0.21  &     $\le$1.5 & (b) \\
Gl809         &   8.54  &   5.43  &   4.92  &   4.60  &           M2.0  &     3597  &   130.00$\pm$3.60  &    0.614$\pm$0.006  &    -0.06  &     $\le$2.8 & (c) \\
Gl812A         &  11.87  &   7.82  &   7.31  &   7.04  &           M3.0  &     3141  &    60.00$\pm$5.10  &    0.431$\pm$0.042  &     0.12  &    $\le$10.0 & (a) \\
Gl815A         &  10.10  &   6.67  &   6.09  &   5.86  &           M3.0  &     3414  &    70.00$\pm$3.30  &    0.608$\pm$0.039  &       --  &    $\le$10.0 & (a) \\
Gl829         &  10.35  &   6.25  &   5.74  &   5.43  &           M4.0  &     3122  &   150.00$\pm$4.80  &    0.371$\pm$0.028  &     0.01  &     $\le$4.0 & (c) \\
Gl849         &  10.42  &   6.51  &   5.90  &   5.56  &           M3.5  &     3195  &   120.00$\pm$3.30  &    0.426$\pm$0.044  &     0.13  &     $\le$2.4 & (c) \\
Gl851         &  10.29  &   6.72  &   6.03  &   5.79  &           M2.0  &     3345  &    80.00$\pm$2.60  &    0.554$\pm$0.053  &       --  &     $\le$2.5 & (b) \\
Gl860A         &   9.59  &   5.57  &   5.04  &   4.76  &           M2.0  &     3152  &    10.00$\pm$4.50  &                --  &      --  &     $\le$3.0 & (c) \\
Gl860B         &  10.30  &   5.57  &   5.04  &   4.76  &           M4.0  &     2919  &    10.00$\pm$4.50  &                --  &      --  &     4.7 & (f) \\
Gl863         &  10.36  &   7.21  &   6.60  &   6.33  &           M0.0  &     3572  &    70.00$\pm$3.70  &    0.541$\pm$0.006  &    -0.20  &     $\le$1.8 & (b) \\
Gl873         &  10.09  &   6.11  &   5.55  &   5.28  &           M4.5  &     3168  &   200.00$\pm$2.60  &    0.315$\pm$0.002  &    -0.20  &     6.9 & (c) \\
Gl875.1         &  11.78  &   7.70  &   7.13  &   6.85  &           M3.5  &     3130  &    70.00$\pm$3.00  &    0.405$\pm$0.042  &     0.11  &    11.0 & (a) \\
Gl876         &  10.17  &   5.93  &   5.35  &   4.98  &           M5.0  &     3172  &   210.00$\pm$5.40  &    0.322$\pm$0.032  &     0.04  &     $\le$2.0 & (f) \\
Gl880         &   8.66  &   5.36  &   4.80  &   4.52  &           M2.0  &       --  &            --  &       --  &       --  &     $\le$2.8 & (c) \\
Gl905         &  12.28  &   6.88  &   6.25  &   5.90  &           M6.0  &     2746  &   320.00$\pm$1.10  &    0.136$\pm$0.008  &       --  &     $\le$1.2 & (c) \\
Gl908         &   8.98  &   5.83  &   5.28  &   5.02  &           M2.0  &     3572  &   177.90$\pm$5.60  &    0.431$\pm$0.046  &    -0.58  &     $\le$3.0 & (c) \\
LHS1885         &  13.65  &   8.59  &   7.99  &   7.66  &           M4.5  &     2733$^{*}$  &    90.00$\pm$2.30  &    0.208$\pm$0.031  &     0.07  &     $\le$3.7 & (f) \\
LHS2065         &     --  &  11.21  &  10.47  &   9.94  &           M9.0  &       --  &       --  &        --  &       --  &    12.0 & (f) \\
LHS2243         &     --  &  11.99  &  11.33  &  10.95  &           M7.5  &     1846$^{**}$  &          --  &       --  &       --  &     7.0 & (f) \\
LHS2397A         &  19.57  &  11.93  &  11.23  &  10.68  &           M8.5  &       --  &    70.00$\pm$2.10  &    --  &       --  &    20.0 & (f) \\
LHS2520         &   12.06 &   7.77   &  7.14   &  6.86  &            M3.5  &   3033    &  --                & --     &       --  &   $\le$2.0   & (c) \\
LHS2632         &     --  &  12.23  &  11.58  &  11.21  &           M7.0  &     1969$^{**}$  &       --  &       --  &       --  &     5.0 & (f) \\
LHS2645         &  18.80  &  12.19  &  11.55  &  11.16  &           M7.0  &     2487  &       --  &            --  &       --  &     9.0 & (f) \\
LHS2924         &  19.58  &  11.99  &  11.23  &  10.69  &           M9.0  &     1665$^{**}$  &    90.00$\pm$1.30  &    0.105$\pm$0.043  &       --  &    11.0 & (f) \\
LHS3376         &  13.46  &   8.74  &   8.26  &   7.93  &           M4.5  &     2922  &   140.00$\pm$5.30  &    0.139$\pm$0.017  &       --  &    14.6 & (f) \\
LP229-17        &   11.75 &   7.18   &  6.53   &  7.10   &          M3.5   &   3265   &   138.00$\pm$40.00 &    0.214$\pm$0.046  &  -0.54    & $\le$2.0 & (c) \\
LP412-31         &  19.21  &  11.76  &  11.07  &  10.64  &           M8.0  &     2191$^{*}$  &          --  &       --  &       --  &    12.0 & (f) \\
LP467-16         &  13.59  &   9.08  &   8.51  &   8.21  &           M5.0  &     2984  &            --  &       --  &       --  &    15.2 & (f) \\
LP731-47         &     --  &     --  &     --  &     --  &           M6.5  &       --  &            --  &       --  &       --  &    11.0 & (f) \\
LP759-25         &     --  &  11.66  &  11.05  &  10.72  &           M5.5  &       --  &            --  &       --  &       --  &    13.0 & (f) \\
LP944-20         &     --  &  10.73  &  10.02  &   9.55  &           M9.0  &     1764$^{*}$  &        --  &       --  &       --  &    31.0 & (f) \\
RG0050-2722         &     --  &  13.61  &  12.98  &  12.54  &           M8.0  &       --  &             --  &       --  &       --  &     4.0 & (f) \\
SDSS011012.22-085627.5 & --    & 14.78 & 14.22 & 13.86 &	M7.0 & -- & -- & -- & -- &	$\le$3.5 & (k) \\	
SDSS021749.99-084409.4 & 19.04 & 14.20 & 13.63 & 13.30 &	M6.0 & -- & -- & -- & -- &	4.0$\pm$0.5 & (k) \\	
SDSS023908.41-072429.3 & --    & 14.81 & 14.21 & 13.76 &	M7.0 & -- & -- & -- & -- &	$\le$3.5 & (k) \\	
SDSS072543.94+382511.4 &   --    & 12.72 & 12.12 & 11.83 &      M7.0 & -- & -- & -- & -- &	$\le$3.5 & (k) \\	
SDSS083231.52+474807.7 &   17.56 & 12.41 & 11.79 & 11.45 & 	M6.0 & -- & -- & -- & -- &	4.0$\pm$0.5 & (k) \\	
SDSS094720.07-002009.5 & --    & 12.26 & 11.63 & 11.35 &	M7.0 & -- & -- & -- & -- &	6.5$\pm$0.5 & (k) \\	
SDSS094738.45+371016.5 &   --    & 12.19 & 11.67 & 11.34 &	M7.0 & -- & -- & -- & -- &	6.0$\pm$0.5 & (k) \\
SDSS110153.86+341017.1 &   --    & 12.78 & 12.20 & 11.99 & 	M7.0 & -- & -- & -- & -- &	$\le$3.5 & (k) \\	
SDSS112036.08+072012.7 &   --    & --    & --    & --    &	M7.0 & -- & -- & -- & -- &	$\le$3.5 & (k) \\	
SDSS125855.13+052034.7 &   --    & 12.64 & 12.06 & 11.69 &	M7.0 & -- & -- & -- & -- &	$\le$3.5 & (k) \\

\hline
\end{tabular}
\medskip
\end{table*}

\begin{table*}
\footnotesize
\begin{tabular}{ccccccccccccc}
\hline
	
SDSS151727.72+335702.4 &   --    & 12.76 & 12.13 & 11.77 & 	M7.0 & -- & -- & -- & -- &	4.5$\pm$0.5 & (k) \\	
SDSS162718.20+353835.7 &   17.56 & 12.35 & 11.73 & 11.35 &	M7.0 & -- & -- & -- & -- &	8.0$\pm$0.5 & (k) \\	
SDSS220334.10+130839.8 &   --    & 14.38 & 13.87 & 13.55 &	M7.0 & -- & -- & -- & -- &	$\le$3.5 & (k) \\	
SDSS225228.50$-$101910.9 & --    & 14.78 & 14.22 & 14.01 &	M7.0 & -- & -- & -- & -- &	$\le$3.5 & (k) \\	
TVLM513-46546         &     --  &  11.87  &  11.18  &  10.71  &           M9.0  &       --  &           --  &       --  &       --  &    60.0 & (f) \\
TVLM868-110639         &     --  &  12.61  &  11.84  &  11.35  &           M9.0  &       --  &           --  &       --  &       --  &    30.0 & (f) \\
VB8         &  16.70  &   9.78  &   9.20  &   8.82  &           M7.0  &     2425  &       --  &         --  &       --  &     9.0 & (f) \\
VB10         &  17.30  &   9.91  &   9.23  &   8.77  &           M8.0  &       --  &       --  &          --  &       --  &     6.5 & (f) \\
YZCMi         &     --  &     --  &     --  &     --  &           M5.5  &       --  &       --  &       --  &       --  &     5.3 & (j) \\

\hline
\end{tabular}

\medskip
Column 12 shows the source of each $v$~sin~$i$ from the literature with references as follows: (a) \citet{stauffer86}; (b) \citet{marcy92a}; (c) \citet{delfosse98}; (d) \citet{glebocki00}; (e) 
\citet{gizis02}; (f) \citet{mohanty03}; (g) \citet{bailer-jones04}; (h) \citet{fuhrmeister04}; (i) \citet{jones05}; (j) \citet{reiners07b}; (k) \citet{west09}

\end{table*}

\section{Conclusions}

We present the initial results from our study of rotation rates for a range of M stars as part of our target selection for a PRVS-like planet search project.  
We observed over 50 M stars with HRS on the HET with the aim of selecting the slowest rotators in order that a near infrared planet search survey, such 
as PRVS, shall have a statistically large sample of M stars where highly precise radial-velocity measurements can be accrued.  Of our sample of
49 suspected single M stars 
between M3-M6.5, we find 36 have $v$~sin~$i$s $\le$~10~km~s$^{-1}$ which will represent good radial-velocity targets.  When we include all literature M stars in the 
optically difficult radial-velocity regime (M3-M9.5) we find this value increases to 124.  

We also confirm the increase of rotational velocities between early to mid M stars.  This change at the fully convective boundary seems to be linked to a change in the topology of the 
magnetic fields between such stars, indicating axisymetric poloidal fields drive a less efficient wind braking mechanism.  Also there is an initial indication that stars with spectral types in 
the range M6.5-M8.5 have a different rotational velocity distribution compared with those of below M6.5, as the distribution appears to flatten off beyond this regime.  Since this is around 
the temperature where dust opacity becomes important 
 ($\sim$2800K), there may be another change in the efficiency of the braking mechanism in such stars which could indicate another magnetic field topology change.  In addition, we also 
show how knowledge of the $v$~sin~$i$ can be used to put a lower, or upper, limit on the age of mid-M dwarf stars, since the rotation-activity relation has a temperature dependent saturation 
level.

We also highlight the rotation-activity relation through emission, or lack thereof, of the
H$\alpha$ line.  There appears a boundary of around 7~km~s$^{-1}$ between stars
with and without H$\alpha$ emission, with the fast rotators almost always
exhibiting such emission, however since the sin~$i$ degeneracy is present it is 
difficult to account for any firm empirical boundaries with small numbers.  We observed the star GJ1253 over two epochs and
found the H$\alpha$ emission had switched on over a period of less than one month.  This may be due to an active region rotating 
in and out of our field of view with a period equal to the rotation period of the star, or due to a flaring event.  

We have also discovered three spectroscopic binary systems and confirmed another.  Both components in the GJ1080 system produce fairly similar profiles, indicating they are of 
similar spectral type, the secondary likely a little cooler and less luminous.  The profiles for GJ3129 and Gl802 are widely separated and both of these exhibit double H$\alpha$ 
emission features.  Gl802 is also a known triple system, but we find evidence for another companion in the system, however given the noise due to blended light from at least 
three separate sources the evidence is weak.  In comparison to these fairly strong profiles the secondary profile in the LHS3080 system is significantly weaker, indicating the companion 
is significantly cooler than its host star.  Finally, we have flagged other M dwarfs where there is some evidence for binary companions in these systems. 


\acknowledgements

J.S.J acknowledges partial support from Centro de Astrof\'\i sica FONDAP 15010003, along with partial support from GEMINI-CONICYT FUND and partial support 
from Comit\'e Mixto ESO-GOBIERNO DE CHILE.  This work also used the research computing facilities at the
Centre for Astrophysics Research, University of Hertfordshire. YP's work was partially supported by the Microcosmophysics program of
National Academy of Sciences and National Space Agency of Ukraine.  The Hobby-Eberly Telescope (HET) is a joint project of the University of Texas at 
Austin, the Pennsylvania State University, Stanford University, Ludwig-Maximilians-Universität München, and Georg-August-Universität Göttingen. The HET 
is named in honor of its principal benefactors, William P. Hobby and Robert E. Eberly.  Our research has made use of the SIMBAD database operated at CDS,
Strasbourg, France.  We also acknowledge the very helpful comments from the anonymous referee.

\bibliographystyle{aa}
\bibliography{refs}

\end{document}